\documentclass[a4paper,prl,twocolumn]{revtex4-1}%

\usepackage{amsmath}
\usepackage{amssymb}
\usepackage{amstext}
\usepackage{color}
\usepackage{graphicx}
\usepackage[space]{grffile}
\usepackage{hyperref}

\begin{document}

\title{Sympathetic cooling of a membrane oscillator in a hybrid mechanical-atomic system}

\author{Andreas~J\"{o}ckel}
\author{Aline~Faber}
\author{Tobias~Kampschulte}
\author{Maria~Korppi}
\author{Matthew~T.~Rakher}
\author{Philipp~Treutlein}
\email[Electronic address: ]{philipp.treutlein@unibas.ch}
\affiliation{Departement Physik, Universit{\"a}t Basel, CH-4056 Basel, Switzerland}

\date{\today}

\maketitle

\textbf{Sympathetic cooling with ultracold atoms \cite{Myatt:1997ct} and atomic ions \cite{Larson:1986ca} enables ultralow temperatures in systems where direct laser or evaporative cooling is not possible. It has so far been limited to the cooling of other microscopic particles, with masses up to $90$ times larger than that of the coolant atom \cite{Offenberg:2008hr}. Here we use ultracold atoms to sympathetically cool the vibrations of a Si$_3$N$_4$ nanomembrane \cite{Thompson:2008dx,Wilson:2009ct}, whose mass exceeds that of the atomic ensemble by a factor of $10^{10}$. The coupling of atomic and membrane vibrations is mediated by laser light over a macroscopic distance \cite{Hammerer:2010fr,Camerer:2011do} and enhanced by placing the membrane in an optical cavity \cite{Vogell:2013fr}. We observe cooling of the membrane vibrations from room temperature to $650\pm 230 \, \textrm{mK}$, exploiting the large atom-membrane cooperativity \cite{Bennett:2014ui} of our hybrid optomechanical system \cite{Hunger:2011eo,Treutlein:2012wa}. Our scheme enables ground-state cooling and quantum control of low-frequency oscillators such as nanomembranes or levitated nanoparticles \cite{Millen:2014dd}, in a regime where purely optomechanical techniques cannot reach the ground state \cite{Vogell:2013fr,Bennett:2014ui}.
}

The control over micro- and nanomechanical oscillators has recently made impressive progress \cite{Aspelmeyer:2012fy}. First experiments cooled high-frequency mechanical oscillators to the quantum ground state \cite{Oconnell:2010br,Teufel:2011jg,Chan:2011dy,Verhagen:2013ei} and demonstrated single-phonon control \cite{Oconnell:2010br} using cryogenic cooling and the techniques of cavity optomechanics \cite{Aspelmeyer:2013vr}. A current challenge is to couple engineered mechanical structures to microscopic quantum systems with good coherence properties, such as atoms \cite{Wang:2006dd,Hunger:2010fr,Camerer:2011do}, solid-state spin systems \cite{Degen:2009kd,Arcizet:2011cg,Kolkowitz:2012iw}, semiconductor quantum dots \cite{Yeo:2013ja}, or superconducting devices \cite{Oconnell:2010br,Pirkkalainen:2013gh}.
The resulting hybrid mechanical systems offer new possibilities for quantum control of mechanical vibrations, precision sensing, and quantum-level signal transduction \cite{Treutlein:2012wa,Hunger:2011eo,Aspelmeyer:2012fy,Aspelmeyer:2013vr}. Ultracold atoms are an attractive choice for hybrid systems because a well-developed toolbox exists for atomic laser cooling and quantum manipulation \cite{Weidemueller:2009}. It has been proposed to use hybrid mechanical-atomic systems for sympathetic cooling \cite{Hammerer:2010fr,Genes:2011jd,Vogell:2013fr}, creating atom-oscillator entanglement \cite{Hammerer:2009gk,Genes:2011jd} and controlling the oscillator on the single-phonon level \cite{Carmele:2013vf}. However, in the experiments reported so far \cite{Wang:2006dd,Hunger:2010fr,Camerer:2011do}, the mechanical-atomic coupling was far too weak to realize any of these possibilities.

Sympathetic cooling of mechanical oscillators with laser-cooled atoms has received particular interest, as it would allow one to relax the constraints of cavity optomechanical and feedback cooling techniques \cite{Vogell:2013fr,Bennett:2014ui}.
 Up to now, sympathetic cooling with atoms and atomic ions has been used to cool other microscopic particles such as different atoms or molecular ions up to the size of proteins \cite{Larson:1986ca,Myatt:1997ct,Offenberg:2008hr}, with applications in cold chemistry and quantum technology \cite{Willitsch:2012gx,Weidemueller:2009}. In these experiments, the coolant and the target species thermalize through short-range collisional or electrostatic interactions in a trap. A large difference in their mass reduces the cooling performance, which has prevented extensions to more massive objects.

In our experiment, we use laser light to interface a nanomechanical membrane oscillator with an ultracold atomic ensemble. 
Radiation-pressure forces couple the membrane vibrations and the atomic motion to the same light field, which mediates a long-distance mechanical atom-membrane interaction \cite{Hammerer:2010fr}.
The membrane is placed inside an optical cavity to enhance the coupling \cite{Vogell:2013fr}, while the atoms are trapped outside the cavity in a separate vacuum chamber. 
This cavity-enhanced modular approach offers both a sizable coupling strength and the possibility to independently manipulate membrane and atoms.

\begin{figure}
\centering
\includegraphics[width=1\columnwidth]{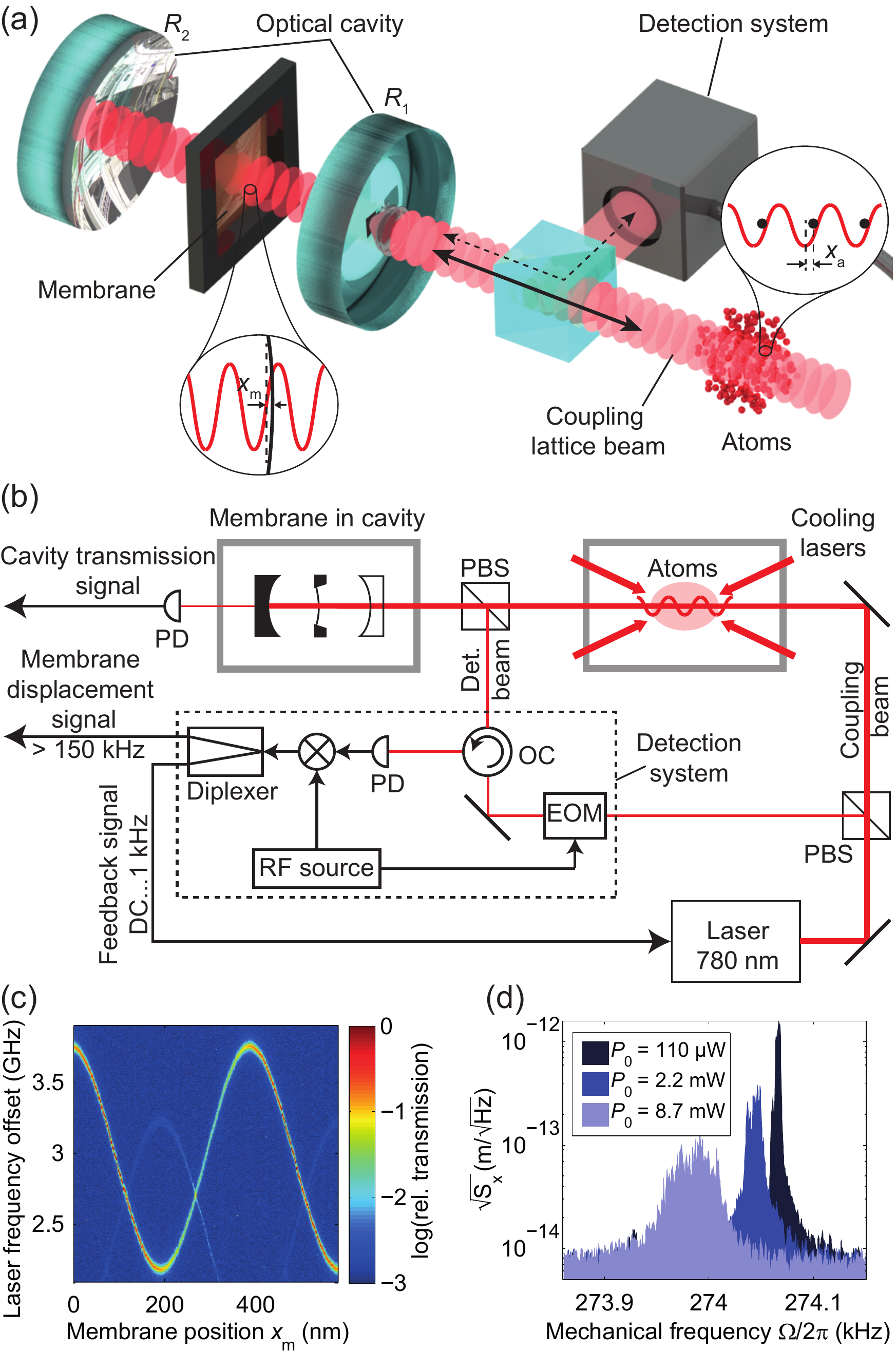}
\caption{\label{fig:paper1} \textbf{Coupling mechanism and schematic of the experiment.} (a) A thin dielectric membrane is placed inside a Fabry-Perot cavity near the slope of the intracavity intensity standing wave. Mirror reflectivities $R_2\approx 1$ and $R_1<R_2$ ensure that the driving laser is reflected and forms an optical lattice for a cloud of ultracold atoms. 
The light couples the membrane vibrations $x_m$ to the atomic motion $x_a$.
The membrane motion is independently read out with a $200\,\mu\textrm{W}$ detection beam.
(b) The membrane-cavity system and the laser-cooled atoms reside in separate vacuum chambers (grey boxes). 
 PBS: polarizing beam splitter, EOM: electro-optic modulator, OC: optical circulator, PD: photodiode. 
(c) Optical transmission spectrum of the cavity showing the cavity resonance frequency $\omega_c$ as a function of membrane position $x_m$ for the fundamental $\textrm{TEM}_\textrm{00}$ (strong signal) and $\textrm{TEM}_\textrm{20}$ (weak signal) mode. 
(d) Mechanical displacement spectra $\sqrt{S_x(\Omega )}$ of the membrane fundamental mode for different lattice laser powers $P_0$ and laser-cavity detuning $\Delta<0$, without atoms in the lattice. 
}
\end{figure}

The coupling mechanism is illustrated in Fig.~\ref{fig:paper1}(a). A thin dielectric membrane acts as mechanical element inside an optical cavity \cite{Thompson:2008dx}, which is resonantly driven by a strong laser beam. The cavity back mirror has a much higher reflectivity than the input mirror, ensuring that the light leaves the cavity through the input port and interferes with the incoming beam to form a standing wave outside the cavity. This standing wave acts as an optical lattice potential for a cloud of ultracold $^{87}$Rb atoms with axial vibration frequency $\Omega_a$ near the bottom of the sinusoidal potential wells \cite{Weidemueller:2009}.
As the fundamental mode of the membrane vibrates with frequency $\Omega_m$ and amplitude $x_m$, it moves back and forth between regions of higher and lower intensity of the intracavity standing wave, thereby modulating the cavity resonance frequency \cite{Thompson:2008dx}. This optomechanical coupling modulates the phase of the laser beam reflected from the cavity, which results in a periodic displacement of the optical lattice potential with amplitude $\propto \mathcal{F} x_m$, enhanced by the cavity finesse $\mathcal{F}\gg 1$. If  $\Omega_a \approx \Omega_m$, this leads to resonant coupling of the membrane vibrations to the atomic center-of-mass motion in the lattice. Conversely, if the atoms oscillate with amplitude $x_a$ 
in the lattice, they redistribute photons in the light field and thereby imprint their motion as a power modulation onto the laser beam driving the cavity. Inside the cavity, this gives rise to a modulation $\propto \mathcal{F} x_a$ of the radiation pressure force experienced by the membrane. 

It was theoretically shown \cite{Vogell:2013fr} that the resulting atom-membrane coupling is described by the Hamiltonian
$H = \hbar g_N \left( \hat b_m^\dagger \hat b_a + \hat b_a^\dagger \hat b_m \right)$, 
where $\hat b_m$ ($\hat b_m^\dagger$) and $\hat b_a$ ($\hat b_a^\dagger$) are the annihilation (creation) operators of the membrane fundamental mode 
and atomic center-of-mass motion, respectively, and 
\begin{equation}
g_N=|r_m|\Omega_a\sqrt{\frac{Nm\Omega_a}{M\Omega_m}}\frac{2\mathcal{F}}{\pi}  \label{eqn:g}
\end{equation}
is the atom-membrane coupling constant for a single vibrational phonon and $N$ atoms of mass $m$ that are identically coupled to the membrane. 
The coupling scales with the optical field reflectivity $r_m$ of the membrane. 
The factor $\sqrt{Nm/M}\ll 1$ describes the mismatch between the mass of the atomic ensemble and the much larger effective mass $M$ of the membrane mode, which reduces the efficiency of the mechanical coupling. 
In previous experiments without a cavity \cite{Camerer:2011do}, this limited the atom-membrane coupling to tens of Hz. Here, the cavity acts as a ``lever'' that partially compensates the mass mismatch by enhancing both the phase shift of the reflected beam and the radiation pressure force acting on the membrane by $\mathcal{F}$.

We exploit this coupling to sympathetically cool the membrane with the atoms.
Strong laser cooling at a rate $\Gamma_a \gg g_N$ cools the atomic ensemble to microkelvin temperatures. 
Due to its interaction with the atoms, the membrane mode is cooled concomitantly with a sympathetic cooling rate \cite{Camerer:2011do,Vogell:2013fr}
\begin{equation}
\Gamma_\textrm{sym} [N,\Omega_a] =  \frac{  g_N^2 \eta^2 t^2 \Gamma_a }{(\Omega_a-\Omega_m)^2 +\left(\Gamma_a/2\right)^2}, \label{eqn:gammaat}
\end{equation}
where we have taken into account the (field) transmittivity $t$ of the optical path between atoms and cavity and the coupling efficiency $\eta$ to the cavity mode (see Appendix).
For a membrane mode coupled with rate $\Gamma_m$ to its thermal environment at temperature $T_\textrm{bath}$ and with $\Gamma_\textrm{sym} \gg \Gamma_m$ to the atoms, whose temperature is negligibly small, the membrane is sympathetically cooled to a temperature $T \simeq T_\textrm{bath} \Gamma_m/\Gamma_\textrm{sym} \ll T_\textrm{bath}$.

%
The experimental setup is sketched in Fig.~\ref{fig:paper1}(b) (see also Appendix). We use a $42\,\textrm{nm}$ thin and $1.5\,\textrm{mm} \times 1.5\,\textrm{mm}$ wide Si$_3$N$_4$ membrane on a Si frame \cite{Wilson:2009ct}. It vibrates like a square drum with a fundamental mechanical mode of frequency $\Omega_m=2\pi\times 274\,\textrm{kHz}$, effective mass $M=140$\,ng, and  energy damping rate $\Gamma_m=0.57\,\textrm{s}^{-1}$. The membrane is semi-transparent for light at our wavelength of $\lambda = 780\,\textrm{nm}$, with $r_m=0.42$ and negligible absorption ($<10^{-5}$). The cavity resonance frequency $\omega_c$ shows a sinusoidal dependence on the membrane position (Fig.~\ref{fig:paper1}(c)), and we typically operate on the slope where $G=-\textrm{d}\omega_c/\textrm{d}x_m$ is largest \cite{Thompson:2008dx}. In our single-sided cavity, $\mathcal{F}$ depends on the membrane position, offering two points of maximal slope with either low $\mathcal{F}=140$ or high $\mathcal{F}=300$. The large cavity linewidth $\kappa \gg \Omega_m, \Omega_a$ ensures that the intracavity field adiabatically follows the membrane and atomic motion. 

The coupling laser beam of frequency $\omega_L$ drives one of the cavity modes with $\eta^2=0.7$ and detuning $\Delta = \omega_L-\omega_c$. We detect the membrane motion with a separate detection beam. Using the Pound-Drever-Hall technique we create a signal whose low-frequency part is used to stabilize $\Delta$. From the high-frequency part we extract the power spectral density of the membrane displacement $S_x(\Omega )$ (Fig.~\ref{fig:paper1}(d)), whose integral is proportional to the temperature $T$ of the membrane mode \cite{Aspelmeyer:2013vr}. To avoid the optomechanical parametric instability \cite{Aspelmeyer:2013vr}, we operate at $\Delta<0$, $|\Delta|\ll \kappa$, resulting in weak cavity optomechanical cooling at a rate $\Gamma_\textrm{opt}$ and the optical spring effect, see Fig.~\ref{fig:paper1}(d), whose known characteristics are used to calibrate our system (see Appendix).

At the position of the atoms, the incoming coupling beam has a power $P_0$, an approximately gaussian intensity profile with beam waist $w_0=284\,\mu$m and $t^2=0.8$. It is red-detuned by $\Delta_\textrm{LA} = -2\pi\times8\,\textrm{GHz}$ from the $F=2\rightarrow F'=3$ transition of the $^{87}$Rb D$_2$ line, creating an attractive optical lattice potential \cite{Weidemueller:2009}. The axial vibration frequency scales as $\Omega_a \propto \sqrt{P_0/|\Delta_\textrm{LA}|}$ and can be adjusted by changing $P_0$ or $\Delta_\textrm{LA}$.
Ultracold atoms are loaded into a 3D-magneto-optical-trap (MOT) overlapping with the optical lattice and further cooled using optical molasses at rate $\Gamma_a=1.0\times 10^4\,\textrm{s}^{-1} $ to a temperature of $40\,\mu \textrm{K}$ and density of $n_a = 8.7\times 10^{15}\,\textrm{atoms}/\textrm{m}^{3}$ in a cloud of $R_a = 3.5\,\textrm{mm}$ radius.

\begin{figure}
\centering
\includegraphics[width=1\columnwidth]{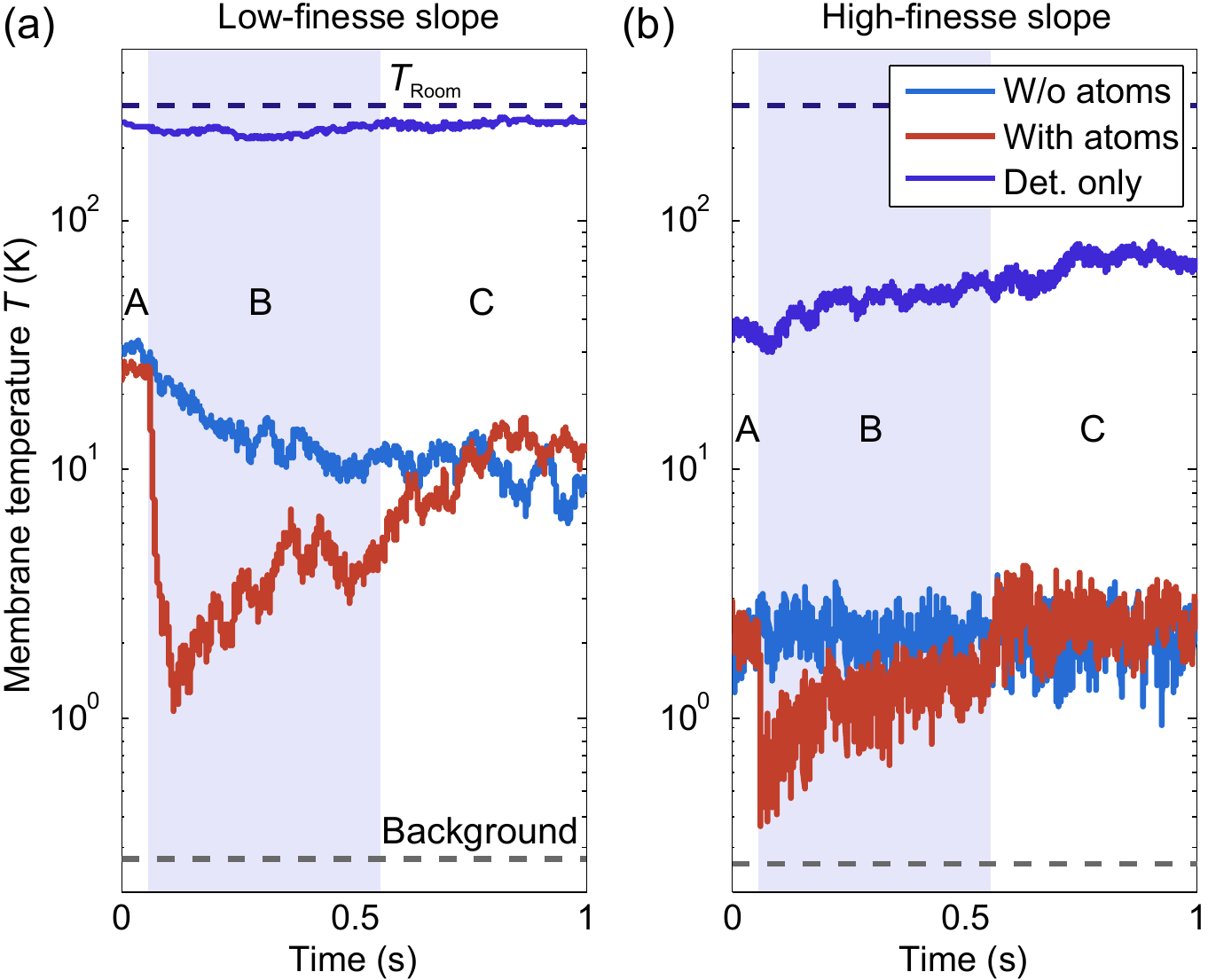}
\caption{\label{fig:paper2}  \textbf{Sympathetic cooling of the membrane.} Membrane temperature $T$ as a function of time in a three-step sequence: (A) atoms not resonant ($P_0=5.5\,\textrm{mW}$); (B) atoms resonant ($P_0=16.5\,\textrm{mW}$); (C) $P_0=16.5\,\textrm{mW}$ but atomic laser cooling switched off. Red curves are recorded with atoms in the lattice, blue curves without atoms. Dark blue curve: readout beam only. Dashed lines: measurement noise floor and room temperature, respectively. Measurements were taken with a spectrum analyzer set to a fixed frequency $\approx \Omega_m$ with bandwidth $\textrm{BW}\gg \Gamma_\textrm{tot}$, and averaged over 20 experimental runs.
(a) Measurement with $\mathcal{F} = 140$, $\Delta = - 0.013 \pm 0.005 \, \kappa$, and $\textrm{BW}=2\pi\times 0.5\,\textrm{kHz}$. 
(b) Measurement with $\mathcal{F} = 300$, $\Delta = - 0.022 \pm 0.002 \, \kappa$, and $\textrm{BW}=2\pi\times 2\,\textrm{kHz}$.
} 
\end{figure}

We first demonstrate sympathetic cooling of the membrane by performing time-resolved measurements of the membrane temperature $T$. A measurement with $\mathcal{F}=140$ is shown in Fig.~\ref{fig:paper2}(a). The three-step sequence starts with $P_0=5.5\,\textrm{mW}$ and a MOT cooled atomic ensemble (configuration A). Cavity optomechanical cooling already cools the membrane from room temperature to 30~K, while the atoms have no effect because $\Omega_a\ll \Omega_m$. We then switch to molasses cooling and $P_0=16.5\,\textrm{mW}$, thereby tuning the atoms into resonance (configuration B). Now a steep decline of the membrane temperature is observed, see the red curve. The initial slope corresponds to the total damping rate of the membrane in the presence of sympathetic and cavity optomechanical cooling, $\Gamma_\textrm{tot}=\Gamma_m + \Gamma_\textrm{sym}+\Gamma_\textrm{opt}$, and we obtain $\Gamma_\textrm{tot}=111 \pm 1\,\mathrm{s}^{-1}$ from a fit to the data. The membrane temperature reaches a minimum of $T_\textrm{sym}=1.5\pm 0.3 \, \textrm{K}$, determined by averaging over 20 experimental runs and a time window of 44~ms, which is about five times the thermalization time $1/\Gamma_\textrm{tot}$. The minimum agrees with the expected value $T_\textrm{sym}=T_\textrm{bath}\Gamma_m / \Gamma_\textrm{tot}$ within one standard deviation. 
The slow increase in $T$ afterwards is caused by atom loss due to the finite molasses lifetime of $0.65\,\textrm{s}$. In configuration C the atomic laser cooling is switched off. 
For comparison, we repeat the experiment without atoms in the lattice (by detuning the MOT lasers), as shown in the blue curve, which sets $\Gamma_\textrm{sym}=0$ so that only cavity optomechanical cooling is present and the temperature equilibrates at $T_\textrm{opt}=T_\textrm{bath} \Gamma_m / (\Gamma_m + \Gamma_\textrm{opt})$. 

We repeat the experiment with $\mathcal{F}=300$, see Fig.~\ref{fig:paper2}(b). This increases both $\Gamma_\textrm{sym}$ and $\Gamma_\textrm{opt}$, resulting in lower temperatures with a minimum of $T_\textrm{sym}=650\pm 230 \, \textrm{mK}$. Here, the uncertainty is mainly due to day-to-day variations of the temperature calibration. 
In all measurements, $\Gamma_\textrm{sym}>\Gamma_\textrm{opt}$ and the atoms provide the dominant cooling effect. 

\begin{figure}
\centering
\includegraphics[width=1\columnwidth]{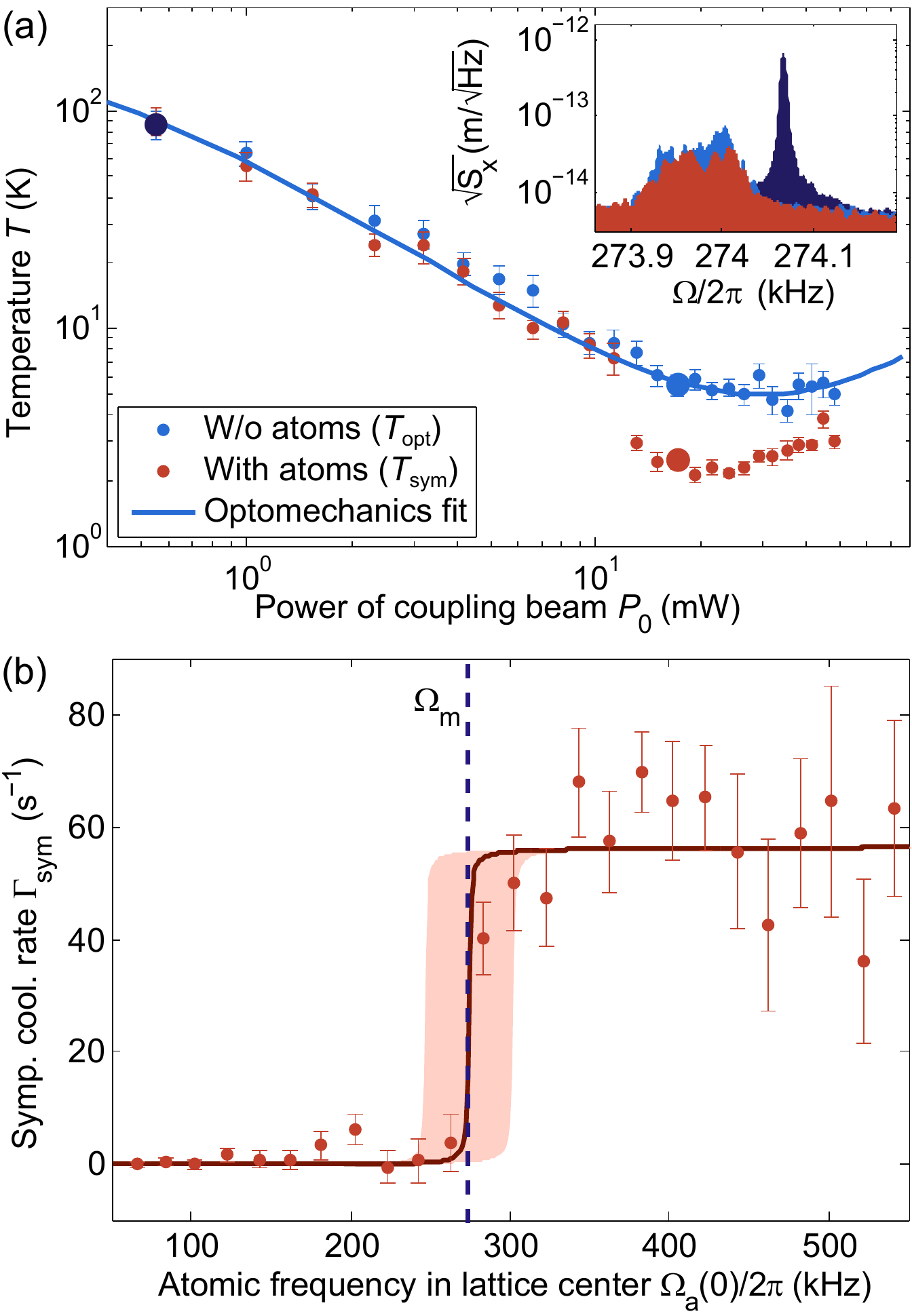}
\caption{\label{fig:paper3}  \textbf{Resonant turn on of sympathetic cooling.} 
(a) Membrane temperature as a function of laser power $P_0$, with atoms in the lattice ($T_\textrm{sym}$) and without atoms ($T_\textrm{opt}$). Blue line: fit with a theory of cavity optomechanical cooling with laser noise, but without atoms. Measurements were performed with $\mathcal{F}=140$, $\Delta = - 0.032 \pm 0.005 \, \kappa$ and averaged over 20 runs. The standard error of the mean (SEM) is shown. The membrane was at a position where $G=0.92\,\max(G)$, decreasing $g_N$ by the same prefactor.
Inset: membrane displacement spectra $\sqrt{S_x(\Omega )}$ corresponding to the big points in the main plot. The spectra are distorted because of fluctuations in $\Delta$ due to a jitter in the membrane position.
(b) Sympathetic cooling rate $\Gamma_\textrm{sym}$ obtained from the data in (a) as a function of atomic frequency in the lattice center, $\Omega_a(0)$. Error bars are calculated from the SEM by error propagation. Red line: model of $\Gamma_\textrm{sym}$ taking the lattice laser profile into account. Shaded red region indicates $\pm 10\%$ uncertainty in $\Omega_a(0)$.  
}
\end{figure}

To study the resonance characteristics of the sympathetic cooling, we measure the membrane temperature with and without atoms as a function of $P_0$, see Fig.~\ref{fig:paper3}(a). Through $P_0$ we effectively adjust the atomic vibration frequency. We now keep $P_0$ constant during the entire sequence and record $S_x ( \Omega )$  in a time window of $380\,\textrm{ms}$ starting when the minimum $T$ is reached. The data without atoms is described well by the theory of cavity optomechanical cooling \cite{Aspelmeyer:2013vr}. We find that laser amplitude and frequency noise limits the achievable cooling factors and extract the effective bath temperature $T_\textrm{bath} (P_0)$ from a fit to the data (see Appendix). 

The data with atoms in Fig.~\ref{fig:paper3}(a) shows that sympathetic cooling turns on abruptly around $P_0 = 14\,\textrm{mW}$. We extract $\Gamma_\textrm{sym}$ from the measured temperatures,
\begin{equation}\label{eqn:gammasymtemp}
\Gamma_\textrm{sym}= \Gamma_m \left( \frac{T_\textrm{bath}}{T_\textrm{sym}}-\frac{T_\textrm{bath}}{T_\textrm{opt}} \right),
\end{equation}
and plot the result in Fig.~\ref{fig:paper3}(b). The horizontal axis shows the atomic frequency in the center of the lattice, calibrated independently by trap spectroscopy (see Appendix). To explain the step-like behavior, we note that the atomic cloud is much larger than the laser beam waist, $R_a \gg w_0$, and that the intensity profile leads to a dependence $\Omega_a(r) = \Omega_a(0) e^{- r^2 / w_0^2}$ on the radial coordinate $r$. 
For small $P_0$, the maximum trap frequency $\Omega_a(0) < \Omega_m$ and none of the atoms are resonant with the membrane, so that $\Gamma_\textrm{sym} \simeq 0$. The sympathetic cooling turns on at the power where $\Omega_a(0) = \Omega_m$, so that atoms in the center of the lattice are resonant. For larger $P_0$, atoms in the center are again off-resonant, but there are always atoms in the wings of the intensity profile for which $\Omega_a(r) = \Omega_m$, so that $\Gamma_\textrm{sym}$ remains constant. 

\begin{figure}
\centering
\includegraphics[width=1\columnwidth]{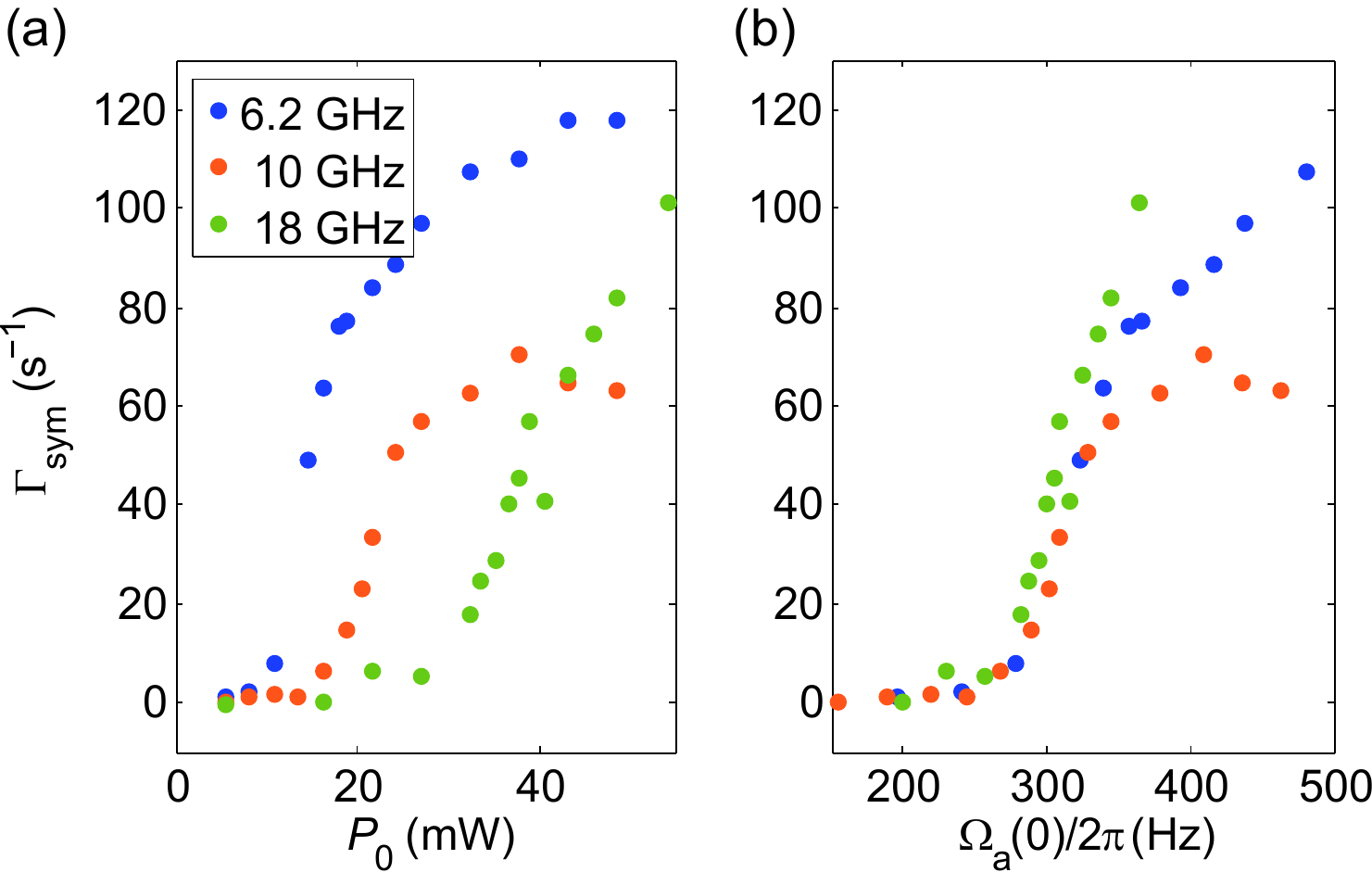}
\caption{\label{fig:paper4} \textbf{Sympathetic cooling rate for different atom-light detunings.}
We measure the sympathetic cooling rate $\Gamma_\text{sym}$ for different atom-light detunings $\Delta_\text{LA}$ and plot the result as a function of (a) power $P_0$  and (b) atomic frequency $\Omega_a(0)$. 
To take this data, we perform membrane ringdown measurements with and without atoms in the lattice and obtain $\Gamma_\text{sym}$ from the difference in decay rates.\cite{Camerer:2011do} We find that $\Gamma_\text{sym}$ starts to rise at different values of $P_0$ for different detunings $\Delta_\text{LA}$ as shown (a). If, on the other hand, we plot the same data as a function of $\Omega_a(0) \propto \sqrt{P_0/|\Delta_\text{LA}|}$, the steps coincide, see (b). This is a further proof that the membrane is cooled by coupling to atomic motion. 
The variation in the overall strength of $\Gamma_\text{sym}$ between datasets is due to a drift of the membrane position inside the cavity.
}
\end{figure}

For a quantitative analysis, we integrate the sympathetic cooling rate over the atomic cloud and obtain a step-like behavior with $\Gamma_\textrm{sym}^\textrm{int} \simeq 4 g_{N_r}^2 \eta^2 t^2 / \Gamma_a$ for $\Omega_a(0) \geq \Omega_m$, where $N_r=\pi^2 n_a w_0^2 R_a \Gamma_a / \Omega_m$ is the number of resonantly coupled atoms (see Appendix). The solid line in Fig.~\ref{fig:paper3}(b) shows a fit with this model using $n_a$ as the only free parameter. From the fit, we obtain $n_a = (4.5 \pm 0.3) \times 10^{15}\,\textrm{atoms}/\textrm{m}^{3}$, which is smaller by a factor $0.52$ than the value measured by absorption imaging at the beginning of the sequence. This is reasonable as atom loss decreases $n_a$ over the duration of the measurement. As a further proof that the membrane is cooled by coupling to atomic motion, we record $\Gamma_\textrm{sym}(P_0)$ for different atom-light detunings $\Delta_\textrm{LA}$ and observe that the step occurs always at the same value of $\Omega_a(0) \propto \sqrt{P_0/|\Delta_\textrm{LA}|}$ (Fig.~\ref{fig:paper4}).

From the data in Fig.~\ref{fig:paper2}(b), we obtain $\Gamma_\textrm{sym}=390\,\textrm{s}^{-1}$, $g_{N_r}=1.3\times 10^3\,\textrm{s}^{-1}$ and $N_r = 9.1\times 10^4$. The mass ratio of the membrane and the resonantly coupled atoms is  $M/(N_r m)=1\times10^{10}$. Still, the light-mediated atom-membrane coupling is efficient and cools the membrane mode a factor $450$ below room temperature. By comparison, previous experiments on the sympathetic cooling of large molecular ions in an ion trap involved mass ratios of up to $\approx 90$, resulting in similar cooling factors and final temperatures \cite{Offenberg:2008hr}.  

In a more general context, our system can be characterized by the atom-membrane cooperativity $C=4 g_N^2 \eta^2 t^2/(\Gamma_a \Gamma_m)$, which quantifies the ability to coherently exchange energy between atoms and membrane \cite{Bennett:2014ui}. It is directly related to the sympathetic cooling rate, and for our data $C=\Gamma_\textrm{sym}/\Gamma_m=680$. For ground-state cooling of the membrane, we require $C>n_\textrm{bath} \simeq k_B T_\textrm{bath}/\hbar \Omega_m$, as has been shown in a fully quantum-mechanical theory of our system that includes effects such as radiation-pressure noise \cite{Vogell:2013fr, Bennett:2014ui}. This regime can be reached with realistic improvements \cite{Vogell:2013fr}, the most important being an increase of atomic density to $n_a=1\times 10^{17}\,\textrm{atoms}/\textrm{m}^{3}$, suppression of technical laser noise, and placing the membrane in a cryogenic environment at $T_\textrm{bath}=4\,$K (see Appendix). 
It is noteworthy that sympathetic ground-state cooling of the membrane is possible in a regime of large $\kappa$ where neither cavity optomechanical nor feedback cooling (cold damping) can reach the ground state \cite{Vogell:2013fr,Bennett:2014ui}. In particular, the resolved-sideband condition $\kappa<\Omega_m$ is not required, making sympathetic cooling an attractive option for low-frequency oscillators such as levitated nanoparticles \cite{Millen:2014dd}.

Our system offers many possibilities for quantum control. 
Because $C>1$, the current setup allows the study of effects analogous to electromagnetically induced transparency \cite{Aspelmeyer:2013vr}.
For the improved parameters mentioned above, switching off atomic laser cooling gives access to the strong-coupling regime \cite{Vogell:2013fr}.
Furthermore, by transducing the membrane vibrations into a polarization change, they can be coupled to the internal atomic state. This allows the creation of Einstein-Podolsky-Rosen entanglement between atoms and membrane without requiring ground-state cooling of the membrane \cite{Hammerer:2009gk}.  
Lastly, Rydberg blockade techniques enable control on the single-phonon level \cite{Carmele:2013vf},
offering exciting opportunities for investigating quantum physics with massive objects.

\section{Acknowledgements} 
We thank Berit Vogell, Klemens Hammerer and Peter Zoller for discussions. This work was supported by the Swiss National Science Foundation through the NCCR Quantum Science and Technology and by the European Union through the project SIQS. M.T.R. acknowledges support from a Marie Curie IIF fellowship.

\section{Appendix}

\subsection{Membrane-cavity system}
We use a commercial Si$_3$N$_4$ square membrane of thickness $d=42\,\textrm{nm}$ and lateral size $l \times l =1.5\,\textrm{mm} \times 1.5\,\textrm{mm}$ supported by a Si frame\cite{Wilson:2009ct}. The fundamental square drum mode of the membrane has a frequency of  $\Omega_m=2\pi\times 274\,\textrm{kHz}$ and an intrinsic mechanical quality factor of $Q_m=\Omega_m/\Gamma_m = 3\times 10^6$ at room temperature, where $\Gamma_m$ is the mechanical energy damping rate. 
The membrane is mounted inside a single-sided Fabry-Perot cavity of length $L=26\,\textrm{mm}$ whose front (back) mirror has an intensity reflectivity of $R_1=96.5$\% ($R_2>99.99$\%). Attocube positioners are used for membrane angular alignment and positioning along the optical axis. 

The optical transmission and reflection spectra of the cavity-membrane system are calculated by solving a system of two coupled cavities, with the membrane as the common mirror\cite{Jayich:2008iz}. The cavity resonance frequencies depend on the membrane position as
\begin{equation}\label{eq:omcmembrane}
\omega_c(x_m)  =  \frac{\omega_\text{FSR}}{\pi} \arccos\left[|r_m| \cos\left(\frac{4\pi x_m}{\lambda}\right)\right]+\omega_0,
\end{equation}
with membrane (field) reflectivity $r_m$, optical wavelength $\lambda = 780$~nm, frequency offset $\omega_0$, cavity free spectral range $\omega_\text{FSR}=\pi c/L=2\pi\times 5.7\,\textrm{GHz}$, and speed of light $c$. 
Using this equation we fit the cavity transmission spectrum (see Fig.~1(c)) with the membrane reflectivity as free parameter, which yields $|r_m|=0.42$. Calculating the expected reflectivity of a thin dielectric plate allows us to determine $d$, assuming a refractive index\cite{Wilson:2009ct} of $2.0$ for Si$_3$N$_4$. 
From the measured $\omega_c(x_m)$ we determine the cavity frequency shift per membrane displacement, $G=-\text{d}\omega_c/\text{d}x_m$. The maxima are given by $G_\text{max}=2 |r_m| \omega_c/L = 2\pi\times 12.2 \,\text{MHz}/\text{nm}$.

In a single-sided cavity, the cavity finesse $\mathcal{F}=\omega_\text{FSR}/\kappa$ depends on the membrane position, which is visible as a change in peak height (color scale) in Fig.~1(c). The observed values of $\mathcal{F}=140\dots 300$ correspond to a cavity intensity decay rate of $\kappa=2\pi\times\left(41 \dots 19\right) \,\textrm{MHz}$. The expected finesse is determined from the calculated transmission spectrum and fit to the data, with $R_1$ as free parameter, assuming $R_2=99.99\%$.
There are two regions with maximal dispersive coupling $|G|$ but different $\mathcal{F}$, corresponding to the light field being mostly in the front sub-cavity (low finesse) or back sub-cavity (high finesse).  Interestingly, in between there is a region with zero dispersive coupling but finite slope $\text{d}\mathcal{F}/\text{d}x_m$, corresponding to a dissipative coupling regime\cite{Elste:2009hf,Weiss:2013gr}.

\subsection{Cavity optomechanical cooling, laser noise and temperature calibration}

The lattice and detection light derives from an external cavity stabilized diode laser at $\lambda=780\,\textrm{nm}$ with a tapered amplifier. The laser is stabilized to a reference cavity to reduce laser frequency noise. The light is split into two paths, one for the lattice beam and one for the membrane detection. Both paths are intensity stabilized with an acousto-optic modulator of the same frequency and coupled to the cavity with orthogonal polarization. Sidebands at $237\,\textrm{MHz}$ are phase-modulated onto the detection beam to create a Pound-Drever-Hall error signal which is used to lock the laser frequency to the cavity resonance and to independently readout the membrane vibrations.

In the absence of atoms in the lattice, the interaction between the membrane vibrations and the intracavity light field still gives rise to cavity optomechanical cooling of the membrane with a cooling rate $\Gamma_\text{opt}$ and a membrane frequency shift $\delta \Omega_m$, the optical spring effect\cite{Aspelmeyer:2013vr}. These phenomena have previously been studied and we use their known dependence on laser power and detuning to calibrate our experiment. 
The cavity optomechanical cooling rate is given by\cite{Aspelmeyer:2013vr}
\[
\Gamma_\text{opt}  =  g_0^2 \bar{n}_c \left( \frac{\kappa}{\frac{\kappa^2}{4}+(\Delta+\Omega_m)^2} - \frac{\kappa}{\frac{\kappa^2}{4}+(\Delta-\Omega_m)^2}\right),\\
\]
where $g_0=G x_{m,0}$ is the single-phonon optomechanical coupling strength, $x_{m,0}=\sqrt{\hbar/2M\Omega_m}$ the membrane zero-point motion and 
\begin{equation}
\bar{n}_c = \frac{\kappa}{\frac{\kappa^2}{4}+\Delta^2} \cdot \frac{P_\text{tot}}{\hbar \omega_c}
\end{equation}
the intracavity photon number. Here, $P_\text{tot}=\eta^2 (t^2 P_0 + P_\text{det})$ is the total laser power coupled to the cavity. The lattice laser power $P_0$ is measured at the position of the atoms, $t^2=0.80$ is the measured transmittance of the optical path between atoms and cavity, and $\eta^2=0.7$ is the coupling efficiency to the cavity TEM$_{00}$ mode, which we estimate from the relative intensity of the TEM$_{00}$ and higher-order transverse modes. Moreover, we have taken into account the small contribution $P_\text{det}$ due to the readout beam.
Cavity optomechanical cooling decreases the membrane temperature to 
$T_\text{opt}=T_\text{bath}\Gamma_m/(\Gamma_m+\Gamma_\text{opt})$.
This simple expression is valid as long as the quantum limits of cavity optomechanical cooling do not yet play a role, which is well satisfied in our experiment. 

The blue datapoints in Fig.~3(a) show measurements of $T_\text{opt}$ for different power levels $P_0$, in the absence of atoms in the lattice.
Experimentally, the temperature of the membrane is determined via the equipartition theorem from the measured variance of the displacement, $k_B T = M \Omega_m^2 \langle x^2 \rangle$, where $\langle x^2 \rangle =  \int_0^{\infty}  \frac{d\Omega}{2\pi} S_x(\Omega)$ is given by the integral of the (single-sided) power spectral density of the mechanical displacement, $S_x(\Omega)$. To calibrate $S_x$ as recorded by the spectrum analyzer, we use the room-temperature thermal motion of the membrane as a reference, see below.

At low power $P_\text{tot}$, the bath temperature $T_\text{bath} \simeq T_0$, which is the temperature of the membrane support (room temperature in our case). At larger $P_\text{tot}$, we observe that laser noise starts to limit the optomechanical cooling\cite{Aspelmeyer:2013vr}. It can be modeled as an effective increase of the bath temperature $T_\text{bath}= T_0 + T_L$, where
\begin{equation}
T_L = \frac{S_\text{F,int}(\Omega_m)+S_\text{F,freq}(\Omega_m)}{ 4 M \Gamma_m k_B }.
\end{equation}
Here, $S_\text{F,int}(\Omega_m)$ and $S_\text{F,freq}(\Omega_m)$ are the force noise power spectral densities (PSD) due to laser intensity and frequency noise, respectively, which are evaluated at $\Omega_m$. For our experiment where $\Omega_m < |\Delta| \ll \kappa$, we find
\begin{eqnarray}
S_\text{F,int}(\Omega_m) & = & \left( \hbar G \bar{n}_c \right)^2 S_I (\Omega_m) \quad \text{and} \\
S_\text{F,freq}(\Omega_m) & = & \left( \hbar G \bar{n}_c \right)^2 \left( \frac{8 \Delta}{\kappa^2} \right)^2 S_{\dot{\phi}} (\Omega_m),
\end{eqnarray}
with $\hbar G \bar{n}_c$ the mean radiation pressure force experienced by the membrane and $S_I (\Omega_m)$ the relative laser intensity noise PSD. The PSD of the frequency noise $S_{\dot{\phi}} (\Omega_m)$ is converted to relative intensity noise by the prefactor $(8\Delta/\kappa^2)^2$.

Since $\bar{n}_c \propto P_\text{tot}$, the cooling rate scales as $\Gamma_\text{opt} \propto P_\text{tot}$ and the laser noise temperature as $T_L \propto P_\text{tot}^2$. To calibrate the parameters of our system, we thus fit a function $T_\text{opt} = \Gamma_m (T_0 + \beta P_\text{tot}^2) / (\Gamma_m + \alpha P_\text{tot})$ to measurements of $T_\text{opt}(P_\text{tot})$, with $\alpha$, $\beta$, and $T_0$ as free parameters, see the solid blue line in Fig.~3(a). From this fit, we obtain the temperature calibration by rescaling $S_x$ so that $T_0 = 295\, \mathrm{K}$. From the parameter $\alpha$ we obtain the detuning $\Delta$, which is difficult to determine by other means in our regime of $|\Delta | \ll \kappa$. All other parameters entering $\alpha$ have been determined independently. In particular, the effective mass $M$ has been obtained from a similar fit to data with a large and known $\Delta$. We find $M=140\,\textrm{ng}$, which is a factor of 2.2 bigger than the calculated effective mass $M_\text{calc}=\rho d l^2 /4$ for a membrane interrogated in the center of the fundamental mode, where $\rho = 2700\,\text{kg}/\text{m}^3$ is the mass density of Si$_3$N$_4$. The difference can be explained by a transverse displacement of the membrane geometric center from the cavity axis, which leads to a smaller displacement (corresponding to larger $M$) at the position where the optical mode overlaps with the membrane.

The parameter $\beta$ allows us to compare the laser noise model with independent measurements of the noise spectra $S_I (\Omega)$ and $S_{\dot{\phi}} (\Omega)$ carried out with an imbalanced Michelson interferometer. We typically measure $S_I (\Omega_m ) = 2.5 \times 10^{-15}/\text{Hz} = -146\,\text{dBc}/\text{Hz}$ and $S_{\dot{\phi}} (\Omega_m )=(2\pi)^2 \times 256\,\text{Hz}^2/\text{Hz}$ at $P_0 = 16.5\,\text{mW}$. At low power we observe a dependence of the relative intensity noise $S_I (\Omega )$ on $P_\text{tot}$. We can neglect this dependence in our model, because at those power levels laser noise does not influence the cooling performance and $T_\text{bath} \simeq T_0$.
The measured noise spectra indicate that both intensity and frequency noise are relevant in our experiment. Comparing the value of $\beta$ calculated from the noise spectra with that obtained from the fit to the optomechanical cooling data, we find agreement to within $14\%$.

We perform such a calibration for the datasets in Fig.~2 and Fig.~3 separately, because the sensitivity of our detection system as well as the laser noise level changes over time. The scatter in the temperature calibration factor is about $35\%$ from day to day. For the lowest measured temperature achieved by sympathetic cooling we have therefore quoted an error bar of $\pm35\%$ as a conservative estimate.

In addition to the optomechanical damping $\Gamma_\text{opt}$ we observe a membrane frequency shift $\delta \Omega_m/P_\text{tot} = -2\pi \times 11\, \text{kHz}/\text{W}$ with laser power $P_\text{tot}$, visible e.g.\ in the inset of Fig.~3(a). The dominant contribution to this shift comes from the optical spring effect\cite{Aspelmeyer:2013vr}, which for $\kappa \gg |\Delta |, \Omega_m$ can be expressed as $\delta \Omega_{m,\text{opt}} = -(\kappa/8\Omega_m) \Gamma_\text{opt}$, and we find $\delta \Omega_{m,\text{opt}}/\delta \Omega_m = 0.89$ for our data.
We explain the residual frequency shift by heating of the membrane due to absorption of laser light, which results in a frequency shift as the thermal expansion reduces the tensile stress in the silicon nitride film. For low-stress silicon nitride membranes, this effect was systematically studied in a recent paper\cite{Jockel:2011ex}. The observed frequency shift for the stoichiometric Si$_3$N$_4$ membranes in our current experiments is reduced by a factor $\approx 10^{4}$ compared to low-stress membranes of the same dimensions and at the same optical power incident on the membrane. 
Using the model of ref.\cite{Jockel:2011ex} but taking into account that the thermal conductivity of stoichiometric Si$_3$N$_4$ is about a factor of 10 higher than that of low-stress silicon nitride\cite{Vogell:2013fr}, we find that at $780\,\text{nm}$, a fraction of $4\times 10^{-6}$ of the optical power is absorbed in a single pass through the membrane. The corresponding  increase of $T_\text{bath}$ in the membrane center is $2.3\,\text{K}$ for $P_0 = 16.5\,\text{mW}$, which has negligible effect on the sympathetic cooling in our experiment.

\subsection{Optical lattice and laser cooling of atoms}

At the position of the atoms, the incoming lattice laser beam has a power $P_0$ and an approximately gaussian intensity profile with a beam waist ($e^{-2}$ radius) of $w_0=284\,\mu$m, so that the peak intensity of the beam is $I_0 = 2 P_0/\pi w_0^2$. The power of the lattice beam that is coupled to the cavity TEM$_{00}$ mode is $P_\text{in} = \eta^2 t^2 P_0$. We observe that the fraction of the light not coupled to the TEM$_{00}$ mode is scattered off the cavity in a divergent spatial mode. If we assume that this light does not find its way back to the atoms, we expect the power of the reflected beam at the position of the atoms, $P_r$, to be reduced by a factor $P_r/P_0 = \eta^2 t^4= 0.45$. An independent measurement yields $P_r/P_0 = 0.51$. 

The interference of the incoming and the reflected beams gives rise to an optical lattice potential of the form 
\begin{equation} \label{eq:Udip}
U_\text{dip}(r,x) = e^{-2 r^2 / w_0^2} \left[ V_d - V_m \sin^2 (k_L x + \phi/2) \right],
\end{equation}
where $k_L = 2\pi /\lambda$ is the wave vector of the light, $V_d = V_0 (1+\eta t^2)^2$ is a constant offset and $V_m = 4 \eta t^2 V_0$ the modulation depth of the potential. The phase shift $\phi$ of the reflected beam with respect to the incoming beam depends on the membrane position, see below.
For a linearly polarized laser beam whose detuning is large compared to the excited-state hyperfine splitting of the $^{87}$Rb $D_2$ line, the single-beam optical dipole potential is
\begin{equation}
V_0 = \frac{\hbar \Gamma^2}{12 \Delta_\text{LA}} \cdot \frac{I_0}{I_s},
\end{equation}
where $I_s = 1.7 \, \text{mW}/\text{cm}^2$ is the saturation intensity of the cycling transition, $\Gamma = 2\pi \times 6.1$\,MHz is the natural linewidth of the $D_2$ line, and the detuning $\Delta_\text{LA} = \omega_L - \omega_{A}$ is defined with respect to the $F=2$ hyperfine component of the ground state\cite{Grimm:2000wx}.
In a harmonic approximation to the bottom of the potential well, the axial vibration frequency of the atoms in the lattice is 
\begin{equation}\label{eq:Omatr} 
\Omega_a(r) = \sqrt{\frac{2 |V_m| k_L^2}{m}} e^{- r^2 / w_0^2}. 
\end{equation}
The dependence on the radial position $r$ in the lattice arises because of the gaussian laser intensity profile. Since $\Omega_a \propto \sqrt{P_0/\Delta_\text{LA}}$, we can tune it by changing the laser beam power $P_0$ and/or the detuning $\Delta_\text{LA}$. 
For $P_0 = 16.5$\,mW and $ \Delta_\text{LA} = -2\pi\times8\,\textrm{GHz}$ as in Fig.~2, we calculate a trap frequency in the center of the lattice of $\Omega_a(0)=2\pi \times 348$\,kHz.

We independently calibrate the atomic trap frequency by applying a weak sinusoidal intensity modulation to the lattice laser beam for a duration of $10$\,ms during which the optical molasses cooling is switched off. This leads to parametric heating of the atom cloud and loss of atoms from the trap. From the observed excitation spectra, we read off a trap frequency of $\Omega_a(0)=2\pi \times 302$\,kHz for the above parameters. This is in reasonable agreement with the calculated value. We attribute the difference to deviations from a gaussian beam profile, which we observe for the beam reflected from the cavity. 
In our analysis of the sympathetic cooling data, we use the measured trap frequencies. 

The spontaneous photon scattering rate in the optical lattice is $\Gamma_\text{sc} = \Gamma U_\text{dip} / \hbar \Delta_\text{LA}$, which amounts to $\Gamma_\text{sc}= 4.0\times 10^4\,\textrm{s}^{-1}$ in the lattice center for $P_0 = 16.5$\,mW and $ \Delta_\text{LA} = -2\pi\times8\,\textrm{GHz}$. 

A magneto-optical trap overlapped with the lattice laser beam is used to prepare $2\times 10^9$ laser-cooled atoms. The atoms are then further cooled using optical molasses to a temperature of  $40\,\mu \textrm{K}$ at a peak density of $n_a \approx 8.7 \times 10^{9}\, \textrm{cm}^{-3}$ as determined from absorption images. The measured molasses lifetime is $0.65\,\textrm{s}$. We measure the atomic laser cooling rate $\Gamma_a=1.0\times 10^4\,\textrm{s}^{-1} $ by recording the atomic temperature decrease after turning on the molasses.

\subsection{Atom-membrane coupling strength and sympathetic cooling rate}

A fully quantized theory of the membrane-light-atom interaction has recently been published\cite{Vogell:2013fr}. It shows that the atom-membrane coupling is described by the Hamiltonian
\begin{equation} \label{eq:Hamilton}
H = \hbar g_N \left( \hat b_m^\dagger \hat b_a + \hat b_a^\dagger \hat b_m \right),
\end{equation}
with the coupling constant $g_N$ given in Eq.~(1) of the paper. In addition, the theory provides expressions for the relevant dissipation mechanisms, the quantum limits of sympathetic cooling, and the conditions for strong coupling. 

The Hamiltonian Eq.~(\ref{eq:Hamilton}) describes the coupling in a lossless system. In our experiment, the cavity incoupling efficiency $\eta<1$ and the optical transmittance between atoms and cavity $t<1$ lead to photon loss. The main consequence for sympathetic cooling is the appearance of a factor $\eta^2 t^2$ in the cooling rate Eq.~(2). More generally, the losses lead to an asymmetry in the coupling, and the theory of cascaded quantum systems is required to describe the dynamics. This has been analyzed in detail for a system where atoms in a lattice are coupled to a membrane without an optical cavity\cite{Hammerer:2010fr,Camerer:2011do}. The losses in the present system can be treated along similar lines.

In the following, we derive the atom-membrane coupling strength including the effect of photon loss, treating the light field classically. This is justified for our parameters where the light acts essentially as a ``spring'' between atoms and membrane. For a single-sided cavity with intensity decay rate $\kappa$ driven by a laser beam with detuning $\Delta = \omega_L - \omega_c$ from cavity resonance, the phase shift of the beam reflected from the cavity with respect to the incoming beam is\cite{Aspelmeyer:2013vr} 
\begin{equation}
\phi = \arctan \left[ \frac{\kappa \Delta}{(\kappa/2)^2-\Delta^2} \right],
\end{equation}
whose derivative with respect to the detuning is
\begin{equation}
\frac{\text{d}\phi}{\text{d}\Delta} =  \frac{\kappa}{(\kappa/2)^2+\Delta^2}.
\end{equation}
For $|\Delta|\ll \kappa$ as in our experiment, we obtain $\phi \simeq 0$ and $\text{d}\phi/\text{d}\Delta \simeq 4/\kappa$.
A small membrane displacement $x_m$ shifts the cavity frequency by $\delta \omega_c = -G x_m$ and thus the phase of the reflected beam by $\delta \phi = -(\text{d}\phi/\text{d}\Delta) \delta \omega_c = (4/ \kappa) G x_m $. This leads to a displacement of the minima of the optical lattice potential by $\delta x =- \delta \phi /2 k_L$, see Eq.~(\ref{eq:Udip}). In a harmonic approximation to the bottom of the lattice potential wells, the resulting force on each atom is $F_a=m \Omega_a^2 \delta x = -(2G/k_L \kappa) m \Omega_a^2 x_m$. The force on the center-of-mass of an ensemble of $N$ atoms is $F_\text{cm}=NF_a$. For a membrane placed at the slope of the intracavity standing wave we have $G=G_\text{max}$ and  find
\begin{equation}
F_\text{cm} = -2 |r_m| \frac{2\mathcal{F}}{\pi} N m \Omega_a^2 x_m = - K x_m, 
\end{equation}
where we have defined a coupling spring constant $K$ which couples the membrane displacement to the center-of-mass motion of the atoms in the lattice\cite{Camerer:2011do,Vogell:2013fr,Bennett:2014ui}.
It is directly connected to the single-phonon coupling constant $g_N = K x_{m,0} x_{a,0}/\hbar$, where $x_{m,0}=\sqrt{\hbar/2M\Omega_m}$ and $x_{a,0}=\sqrt{\hbar/2Nm\Omega_a}$ are the quantum mechanical zero-point amplitudes of the membrane and atomic center-of-mass motion, respectively.

Conversely, an atom displaced by $x_a$ from the bottom of the potential well experiences a restoring optical dipole force $F_d = -m \Omega_a^2 x_a$. On a microscopic level, this force is due to absorption and stimulated emission, leading to a redistribution of photons between the incoming and reflected laser beams which form the lattice potential\cite{Morrow:2002vp}. Each absorption-emission event between the counterpropagating beams imparts a momentum kick of $\pm 2 \hbar k_L$ to the atom. For $N$ atoms in the lattice, the corresponding photon redistribution rate is $\dot n_p = N F_d / (2 \hbar k_L)$, which leads to a power modulation of the laser beam after it has passed through the atomic ensemble of $\delta P_0 = \dot n_p \hbar \omega_L = (c/2) N F_d$, where $c$ is the speed of light.
The power modulation of the coupling beam leads to a modulation of the mean photon number $\bar{n}_c$ inside the cavity. For a single-sided cavity, we have
\begin{equation} 
\bar{n}_c = \frac{\kappa}{\frac{\kappa^2}{4}+\Delta^2} \cdot \frac{P_\text{in}}{\hbar \omega_c},
\end{equation}
with $P_\text{in} = \eta^2 t^2 P_0$ the input power to the cavity. For $|\Delta|\ll \kappa$ we obtain a modulation $\delta \bar{n}_c \simeq (4/\kappa) \eta^2 t^2 (\delta P_0 /\hbar \omega_c)$.
In an optomechanical system\cite{Aspelmeyer:2013vr}, the mean radiation-pressure force experienced by the mechanical element is $F_\text{rad}= \hbar G \bar{n}_c$. The motion of the atoms in the lattice thus leads to a modulation of the radiation pressure force on the membrane of $\delta F_\text{rad} = \hbar G \delta \bar{n}_c = - \eta^2 t^2 (2 G/k_L \kappa) N m \Omega_a^2 x_a$, which for $G=G_\text{max}$ is
\begin{equation}
\delta F_\text{rad} = - \eta^2 t^2 2 |r_m| \frac{2\mathcal{F}}{\pi} N m \Omega_a^2 x_a = - \eta^2 t^2 K x_a.
\end{equation}
We thus find that for a comparable displacement, the force on the membrane is smaller by a factor $\eta^2 t^2$ than the corresponding force on the atoms\cite{Camerer:2011do}.

The membrane vibrations and the atomic center-of-mass motion can be described as harmonic oscillators coupled through $F_\text{cm}$ and $\delta F_\text{rad}$, with equations of motion\cite{Camerer:2011do,Bennett:2014ui}
\begin{eqnarray*}
N m \ddot{x}_a &=& -\Gamma_a N m \dot{x}_a - N m \Omega_a^2 x_a - K x_m , \\
M \ddot{x}_m &=& -\Gamma_m M \dot{x}_m - M \Omega_m^2 x_m - \eta^2t^2 K x_a + F_\text{th},
\end{eqnarray*}
where $F_\text{th}$ describes the fluctuating thermal force due to the coupling of the membrane to the environment at temperature $T_\text{bath}$. Since the atomic temperature is negligibly small in our experiment, we have suppressed the corresponding term. Fourier transforming the equations of motion we obtain 
\begin{equation}\label{eq:motionfourier}
\begin{split}
\tilde{x}_a(\Omega) &= \chi_a(\Omega) \left[ -K \tilde{x}_m(\Omega) \right] \quad \text{and} \\
\tilde{x}_m(\Omega) &= \chi_m(\Omega) \left[ \tilde{F}_\text{th}-\eta^2t^2 K \tilde{x}_a(\Omega)\right],
\end{split}
\end{equation} 
with the mechanical susceptibilities
\begin{eqnarray*}
\chi_a(\Omega) &\simeq&  \left[ 2Nm\Omega_a (\Omega_a - \Omega - i \Gamma_a/2) \right]^{-1} \quad \text{and} \\
\chi_m(\Omega) &\simeq&  \left[ 2M\Omega_m (\Omega_m - \Omega - i \Gamma_m/2) \right]^{-1},
\end{eqnarray*}
where we made a Lorentzian approximation valid for $\Omega_a \gg \Gamma_a$ and $\Omega_m \gg \Gamma_m$. Eliminating $\tilde{x}_a$ in Eqs.~(\ref{eq:motionfourier}) we obtain for the membrane amplitude
\begin{equation}\label{eq:xmtilde}
\begin{split}
\tilde{x}_m(\Omega) &= \frac{\tilde{F}_\text{th}}{\chi_m^{-1}(\Omega) - \eta^2t^2 K^2 \chi_a(\Omega)} \\
 &= \frac{\tilde{F}_\text{th}}{ 2 M \Omega_m \left(  \Omega_m - \Omega - i \frac{\Gamma_m}{2} 
- \frac{\eta^2 t^2 g_N^2}{\Omega_a - \Omega - i \Gamma_a/2 }  \right)  }.
\end{split}
\end{equation}
For $\Gamma_a \gg g_N, \Gamma_m$ as in our experiment, we can replace $\Omega \rightarrow \Omega_m$ in the last term in the denominator and obtain
\begin{equation}
\begin{split}
\tilde{x}_m(\Omega) &= \frac{\tilde{F}_\text{th}}{ 2 M \Omega_m \left(  \Omega_m - \delta\Omega_m - \Omega - i \frac{\Gamma_m + \Gamma_\text{sym}}{2}  \right) } \\
 &=  \chi_m^\prime (\Omega) \tilde{F}_\text{th},
 \end{split}
\end{equation}
where $\Gamma_\text{sym}$ is the sympathetic cooling rate given in Eq.~(2) of the paper, $\delta\Omega_m = (\Omega_a - \Omega_m)\Gamma_\text{sym}/\Gamma_a $ is a frequency shift of the membrane resonance, and $\chi^\prime (\Omega)$ is the effective susceptibility of the membrane coupled to the atoms. 
The additional damping introduced by $\Gamma_\text{sym}$ decreases the response of the membrane to the thermal force $\tilde{F}_\text{th}$ and thus leads to cooling.
The membrane displacement spectrum is given by $S_x(\Omega) = | \chi^\prime (\Omega) |^2 S_\text{F,th}$, where $S_\text{F,th}=4 M \Gamma_m k_B T_\text{bath}$ is the (single-sided) thermal force noise PSD. Integrating the spectrum, we obtain a membrane steady-state temperature of 
$T=T_\text{bath}\Gamma_m/(\Gamma_m+\Gamma_\text{sym})$.
We note that in the limit $g_N \gg \Gamma_a, \Gamma_m$, Eq.~(\ref{eq:xmtilde}) describes a hybridization of the atom and membrane oscillators into normal modes\cite{Bennett:2014ui}.
\\
\\
\subsection{Ensemble-integrated sympathetic cooling rate}
In our experiment, the cloud of molasses atoms has a radius $R_a$ that is much larger than the waist $w_0$ of the coupling laser beam. Atoms at different radial positions $r$ in the lattice experience different axial vibration frequencies, $\Omega_a(r) = \Omega_a(0) e^{- r^2 / w_0^2}$, see Eq.~(\ref{eq:Omatr}). The dependence of $\Omega_a$ on the longitudinal position in the lattice is negligible. From absorption images, we find that the atomic number density $n_a$ in the molasses is approximately constant over the lattice profile. To quantitatively model the data in Fig.~3(b), we integrate the sympathetic cooling rate over the lattice laser beam profile,
\begin{equation}
\Gamma_\text{sym}^\text{int} = 2 R_a n_a \int_0^{R_a} \text{d}r \, 2\pi r \Gamma_\text{sym}[N=1,\Omega_a(r)].
\end{equation}
Converting this to an integral over frequency we have
\begin{equation}
\Gamma_\text{sym}^\text{int} = N_\text{lat} \int_{\Omega_a(R_a)}^{\Omega_a(0)} \text{d}\Omega_a \frac{\Gamma_\text{sym}[N=1,\Omega_a]}{\Omega_a},
\end{equation}
where $N_\text{lat}=2 R_a \pi w_0^2 n_a$ is the number of atoms in the lattice volume. Using Eq.~(2) and (1) we obtain
\begin{equation}
\Gamma_\text{sym}^\text{int} = B \int_{\Omega_a(R_a)}^{\Omega_a(0)} \text{d}\Omega_a \frac{\Omega_a^2}{\left( \Omega_a-\Omega_m \right)^2 + \left( \Gamma_a/2\right)^2},
\end{equation}
where $B=  |r_m|^2 \tfrac{m N_\text{lat}}{M}(\tfrac{2\mathcal{F}}{\pi})^2 \eta^2 t^2 \tfrac{\Gamma_a}{\Omega_m}$. Since $R_a \gg w_0$ we take the lower integration limit $\Omega_a(R_a) \rightarrow 0$ and find
\begin{widetext} 
\begin{equation}\label{eqn:gammasymint}
\begin{split}
 \Gamma_\text{sym}^\text{int} = \frac{4 g_{N_r}^2 \eta^2 t^2}{\Gamma_a\pi}
 & \left\{ \left( 1-\frac{\Gamma_a^2}{4\Omega_m^2} \right) \left( \arctan\left[\frac{2\Omega_m}{\Gamma_a}\right]+\arctan\left[\frac{2(\Omega_a(0)-\Omega_m)}{\Gamma_a}\right]\right) \right. \\
 &+ \left. \frac{\Gamma_a}{2\Omega_m^2} \left( \Omega_a(0)+\Omega_m \ln\left[\frac{\Gamma_a^2+4(\Omega_a(0)-\Omega_m)^2}{\Gamma_a^2+4\Omega_m^2}\right] \right) \right\}.
\end{split}
\end{equation}
\end{widetext}
Here, $N_r= N_\text{lat} (\pi \Gamma_a/2 \Omega_m)=\pi^2 R_a w_0^2 n_a \Gamma_a / \Omega_m $ is the number of resonantly coupled atoms and $g_{N_r}=|r_m|\Omega_m\sqrt{\frac{N_r m}{M}}\frac{2\mathcal{F}}{\pi}$ the corresponding coupling constant. 
For $\Gamma_a\ll\Omega_m$ as in our experiment we can approximate Eq.~(\ref{eqn:gammasymint}) as
\begin{equation}
\begin{split}
 \Gamma_\text{sym}^\text{int} \simeq \frac{4 g_{N_r}^2 \eta^2 t^2}{\Gamma_a\pi} 
 & \left( \arctan\left[\frac{2\Omega_m}{\Gamma_a}\right] \right. \\
 &+ \left. \arctan\left[\frac{2(\Omega_a(0)-\Omega_m)}{\Gamma_a}\right]\right),
\end{split}
\end{equation}
which is a step-like function with step width $\Gamma_a$ and step height $4 g_{N_r}^2 \eta^2 t^2 / \Gamma_a$.  To analyze our measurement we fit the complete function in Eq.~(\ref{eqn:gammasymint}) to the data with $n_a$ as the only free parameter.

\subsection{Parameters required for ground-state cooling}

In our experiment, a minimum phonon number of $n_\text{ss}=5\times10^4$ was reached cooling down from an effective thermal bath occupation of $n_\text{bath}=5\times10^7$. In this section we provide parameters of an improved experimental setup that can sympathetically cool the membrane mode to the vibrational ground state ($n_\text{ss}<1$). The estimate is based on a fully quantum mechanical description of our system\cite{Vogell:2013fr} that includes dissipation mechanisms such as radiation pressure noise on membrane and atoms and thermal heating of the membrane by laser absorption. In addition, we consider technical laser noise and the spatial variation of the cooling rate due to the laser intensity profile. 

In the improved setup, the membrane is cryogenically pre-cooled to $T_0=4\,\text{K}$. The mechanical quality factor can be increased using higher vibrational modes of the membrane\cite{Chakram:2014ea}, and we consider the $(4,4)$-mode with $\Omega_m=2\pi \times 1.1\,\text{MHz}$ and $Q_m=4\times 10^7$. Using a shorter cavity ($L=1\,\text{mm}$) of moderate finesse $\mathcal{F}=10^3$ increases $\kappa$ and thus suppresses the effect of laser frequency noise. Regarding intensity noise, we point out that shot-noise-limited operation has been achieved in a membrane optomechanical setup at two orders of magnitude higher intracavity photon number than required here\cite{Purdy:2013cb}. Improving the alignment, we have $M=63.5\,\text{ng}$, and $t^2=\eta^2=0.9$ seem feasible. 

A blue-detuned coupling lattice in combination with a far-detuned transverse 2D lattice suppresses light-assisted collisions of the atoms, enabling a smaller atom-light detuning $\Delta_\text{LA}=2\pi\times 0.5\,\text{GHz}$ and a higher atomic density $n_a=1 \times 10^{17}\,\text{atoms}/\textrm{m}^{3}$. Raman sideband cooling has been used to cool large ensembles of atoms to the vibrational ground state of a lattice at comparable densities\cite{Kerman:2000th}. We assume a laser cooling rate $\Gamma_a =4\times 10^4\,\textrm{s}^{-1}$, $P_0 = 0.5\,\text{mW}$, and $w_0=70\,\mu\text{m}$. With these parameters, the resonant atom number is $N_r = 1\times 10^5$, the atom-membrane cooperativity $C=4\times 10^5$ and it is possible to cool the membrane mode to a steady-state phonon number $n_\text{ss}=0.75$, taking into account thermal heating of the membrane by laser absorption\cite{Vogell:2013fr}.

Recently it was shown\cite{Bennett:2014ui} that sympathetic cooling with atoms can provide ground-state cooling of the membrane in a regime where cavity optomechanical cooling cannot reach the ground state because the system operates in the non-resolved sideband limit $\Omega_m \ll \kappa$, and feedback cooling (cold damping) cannot reach it either because the optomechanical cooperativity $c_m = 4 g_0^2 \bar{n}_c /(\Gamma_m \kappa) < n_\text{bath}/8$. Sympathetic cooling with atoms thus provides new opportunities for ground-state cooling of mechanical oscillators in the bad-cavity regime where $\kappa$ is large. Finally, we note that by switching off the atomic laser cooling ($\Gamma_a = 0$) the above parameters give access to the strong coupling regime where $g_N$ is larger than the decoherence rates of both the atoms and the membrane\cite{Vogell:2013fr}.

\bibliography{membranecooling}

\begin{thebibliography}{39}%
\makeatletter
\providecommand \@ifxundefined [1]{%
 \@ifx{#1\undefined}
}%
\providecommand \@ifnum [1]{%
 \ifnum #1\expandafter \@firstoftwo
 \else \expandafter \@secondoftwo
 \fi
}%
\providecommand \@ifx [1]{%
 \ifx #1\expandafter \@firstoftwo
 \else \expandafter \@secondoftwo
 \fi
}%
\providecommand \natexlab [1]{#1}%
\providecommand \enquote  [1]{``#1''}%
\providecommand \bibnamefont  [1]{#1}%
\providecommand \bibfnamefont [1]{#1}%
\providecommand \citenamefont [1]{#1}%
\providecommand \href@noop [0]{\@secondoftwo}%
\providecommand \href [0]{\begingroup \@sanitize@url \@href}%
\providecommand \@href[1]{\@@startlink{#1}\@@href}%
\providecommand \@@href[1]{\endgroup#1\@@endlink}%
\providecommand \@sanitize@url [0]{\catcode `\\12\catcode `\$12\catcode
  `\&12\catcode `\#12\catcode `\^12\catcode `\_12\catcode `\%12\relax}%
\providecommand \@@startlink[1]{}%
\providecommand \@@endlink[0]{}%
\providecommand \url  [0]{\begingroup\@sanitize@url \@url }%
\providecommand \@url [1]{\endgroup\@href {#1}{\urlprefix }}%
\providecommand \urlprefix  [0]{URL }%
\providecommand \Eprint [0]{\href }%
\providecommand \doibase [0]{http://dx.doi.org/}%
\providecommand \selectlanguage [0]{\@gobble}%
\providecommand \bibinfo  [0]{\@secondoftwo}%
\providecommand \bibfield  [0]{\@secondoftwo}%
\providecommand \translation [1]{[#1]}%
\providecommand \BibitemOpen [0]{}%
\providecommand \bibitemStop [0]{}%
\providecommand \bibitemNoStop [0]{.\EOS\space}%
\providecommand \EOS [0]{\spacefactor3000\relax}%
\providecommand \BibitemShut  [1]{\csname bibitem#1\endcsname}%
\let\auto@bib@innerbib\@empty
\bibitem [{\citenamefont {Myatt}\ \emph {et~al.}(1997)\citenamefont {Myatt},
  \citenamefont {Burt}, \citenamefont {Ghrist}, \citenamefont {Cornell},\ and\
  \citenamefont {Wieman}}]{Myatt:1997ct}%
  \BibitemOpen
  \bibfield  {author} {\bibinfo {author} {\bibfnamefont {C.}~\bibnamefont
  {Myatt}}, \bibinfo {author} {\bibfnamefont {E.}~\bibnamefont {Burt}},
  \bibinfo {author} {\bibfnamefont {R.}~\bibnamefont {Ghrist}}, \bibinfo
  {author} {\bibfnamefont {E.}~\bibnamefont {Cornell}}, \ and\ \bibinfo
  {author} {\bibfnamefont {C.}~\bibnamefont {Wieman}},\ }\href {\doibase
  10.1103/PhysRevLett.78.586} {\bibfield  {journal} {\bibinfo  {journal}
  {Physical Review Letters}\ }\textbf {\bibinfo {volume} {78}},\ \bibinfo
  {pages} {586} (\bibinfo {year} {1997})}\BibitemShut {NoStop}%
\bibitem [{\citenamefont {Larson}\ \emph {et~al.}(1986)\citenamefont {Larson},
  \citenamefont {Bergquist}, \citenamefont {Bollinger}, \citenamefont {Itano},\
  and\ \citenamefont {Wineland}}]{Larson:1986ca}%
  \BibitemOpen
  \bibfield  {author} {\bibinfo {author} {\bibfnamefont {D.}~\bibnamefont
  {Larson}}, \bibinfo {author} {\bibfnamefont {J.}~\bibnamefont {Bergquist}},
  \bibinfo {author} {\bibfnamefont {J.}~\bibnamefont {Bollinger}}, \bibinfo
  {author} {\bibfnamefont {W.}~\bibnamefont {Itano}}, \ and\ \bibinfo {author}
  {\bibfnamefont {D.}~\bibnamefont {Wineland}},\ }\href {\doibase
  10.1103/PhysRevLett.57.70} {\bibfield  {journal} {\bibinfo  {journal}
  {Physical Review Letters}\ }\textbf {\bibinfo {volume} {57}},\ \bibinfo
  {pages} {70} (\bibinfo {year} {1986})}\BibitemShut {NoStop}%
\bibitem [{\citenamefont {Offenberg}\ \emph {et~al.}(2008)\citenamefont
  {Offenberg}, \citenamefont {Zhang}, \citenamefont {Wellers}, \citenamefont
  {Roth},\ and\ \citenamefont {Schiller}}]{Offenberg:2008hr}%
  \BibitemOpen
  \bibfield  {author} {\bibinfo {author} {\bibfnamefont {D.}~\bibnamefont
  {Offenberg}}, \bibinfo {author} {\bibfnamefont {C.}~\bibnamefont {Zhang}},
  \bibinfo {author} {\bibfnamefont {C.}~\bibnamefont {Wellers}}, \bibinfo
  {author} {\bibfnamefont {B.}~\bibnamefont {Roth}}, \ and\ \bibinfo {author}
  {\bibfnamefont {S.}~\bibnamefont {Schiller}},\ }\href {\doibase
  10.1103/PhysRevA.78.061401} {\bibfield  {journal} {\bibinfo  {journal}
  {Physical Review A}\ }\textbf {\bibinfo {volume} {78}},\ \bibinfo {pages}
  {061401} (\bibinfo {year} {2008})}\BibitemShut {NoStop}%
\bibitem [{\citenamefont {Thompson}\ \emph {et~al.}(2008)\citenamefont
  {Thompson}, \citenamefont {Zwickl}, \citenamefont {Jayich}, \citenamefont
  {Marquardt}, \citenamefont {Girvin},\ and\ \citenamefont
  {Harris}}]{Thompson:2008dx}%
  \BibitemOpen
  \bibfield  {author} {\bibinfo {author} {\bibfnamefont {J.~D.}\ \bibnamefont
  {Thompson}}, \bibinfo {author} {\bibfnamefont {B.~M.}\ \bibnamefont
  {Zwickl}}, \bibinfo {author} {\bibfnamefont {A.~M.}\ \bibnamefont {Jayich}},
  \bibinfo {author} {\bibfnamefont {F.}~\bibnamefont {Marquardt}}, \bibinfo
  {author} {\bibfnamefont {S.~M.}\ \bibnamefont {Girvin}}, \ and\ \bibinfo
  {author} {\bibfnamefont {J.~G.~E.}\ \bibnamefont {Harris}},\ }\href {\doibase
  10.1038/nature06715} {\bibfield  {journal} {\bibinfo  {journal} {Nature}\
  }\textbf {\bibinfo {volume} {452}},\ \bibinfo {pages} {72} (\bibinfo {year}
  {2008})}\BibitemShut {NoStop}%
\bibitem [{\citenamefont {Wilson}\ \emph {et~al.}(2009)\citenamefont {Wilson},
  \citenamefont {Regal}, \citenamefont {Papp},\ and\ \citenamefont
  {Kimble}}]{Wilson:2009ct}%
  \BibitemOpen
  \bibfield  {author} {\bibinfo {author} {\bibfnamefont {D.~J.}\ \bibnamefont
  {Wilson}}, \bibinfo {author} {\bibfnamefont {C.~A.}\ \bibnamefont {Regal}},
  \bibinfo {author} {\bibfnamefont {S.~B.}\ \bibnamefont {Papp}}, \ and\
  \bibinfo {author} {\bibfnamefont {H.~J.}\ \bibnamefont {Kimble}},\ }\href
  {\doibase 10.1103/PhysRevLett.103.207204} {\bibfield  {journal} {\bibinfo
  {journal} {Physical Review Letters}\ }\textbf {\bibinfo {volume} {103}},\
  \bibinfo {pages} {207204} (\bibinfo {year} {2009})}\BibitemShut {NoStop}%
\bibitem [{\citenamefont {Hammerer}\ \emph {et~al.}(2010)\citenamefont
  {Hammerer}, \citenamefont {Stannigel}, \citenamefont {Genes}, \citenamefont
  {Zoller}, \citenamefont {Treutlein}, \citenamefont {Camerer}, \citenamefont
  {Hunger},\ and\ \citenamefont {H{\"a}nsch}}]{Hammerer:2010fr}%
  \BibitemOpen
  \bibfield  {author} {\bibinfo {author} {\bibfnamefont {K.}~\bibnamefont
  {Hammerer}}, \bibinfo {author} {\bibfnamefont {K.}~\bibnamefont {Stannigel}},
  \bibinfo {author} {\bibfnamefont {C.}~\bibnamefont {Genes}}, \bibinfo
  {author} {\bibfnamefont {P.}~\bibnamefont {Zoller}}, \bibinfo {author}
  {\bibfnamefont {P.}~\bibnamefont {Treutlein}}, \bibinfo {author}
  {\bibfnamefont {S.}~\bibnamefont {Camerer}}, \bibinfo {author} {\bibfnamefont
  {D.}~\bibnamefont {Hunger}}, \ and\ \bibinfo {author} {\bibfnamefont {T.~W.}\
  \bibnamefont {H{\"a}nsch}},\ }\href {\doibase 10.1103/PhysRevA.82.021803}
  {\bibfield  {journal} {\bibinfo  {journal} {Physical Review A}\ }\textbf
  {\bibinfo {volume} {82}},\ \bibinfo {pages} {021803} (\bibinfo {year}
  {2010})}\BibitemShut {NoStop}%
\bibitem [{\citenamefont {Camerer}\ \emph {et~al.}(2011)\citenamefont
  {Camerer}, \citenamefont {Korppi}, \citenamefont {J{\"o}ckel}, \citenamefont
  {Hunger}, \citenamefont {H{\"a}nsch},\ and\ \citenamefont
  {Treutlein}}]{Camerer:2011do}%
  \BibitemOpen
  \bibfield  {author} {\bibinfo {author} {\bibfnamefont {S.}~\bibnamefont
  {Camerer}}, \bibinfo {author} {\bibfnamefont {M.}~\bibnamefont {Korppi}},
  \bibinfo {author} {\bibfnamefont {A.}~\bibnamefont {J{\"o}ckel}}, \bibinfo
  {author} {\bibfnamefont {D.}~\bibnamefont {Hunger}}, \bibinfo {author}
  {\bibfnamefont {T.~W.}\ \bibnamefont {H{\"a}nsch}}, \ and\ \bibinfo {author}
  {\bibfnamefont {P.}~\bibnamefont {Treutlein}},\ }\href {\doibase
  10.1103/PhysRevLett.107.223001} {\bibfield  {journal} {\bibinfo  {journal}
  {Physical Review Letters}\ }\textbf {\bibinfo {volume} {107}},\ \bibinfo
  {pages} {223001} (\bibinfo {year} {2011})}\BibitemShut {NoStop}%
\bibitem [{\citenamefont {Vogell}\ \emph {et~al.}(2013)\citenamefont {Vogell},
  \citenamefont {Stannigel}, \citenamefont {Zoller}, \citenamefont {Hammerer},
  \citenamefont {Rakher}, \citenamefont {Korppi}, \citenamefont {J{\"o}ckel},\
  and\ \citenamefont {Treutlein}}]{Vogell:2013fr}%
  \BibitemOpen
  \bibfield  {author} {\bibinfo {author} {\bibfnamefont {B.}~\bibnamefont
  {Vogell}}, \bibinfo {author} {\bibfnamefont {K.}~\bibnamefont {Stannigel}},
  \bibinfo {author} {\bibfnamefont {P.}~\bibnamefont {Zoller}}, \bibinfo
  {author} {\bibfnamefont {K.}~\bibnamefont {Hammerer}}, \bibinfo {author}
  {\bibfnamefont {M.~T.}\ \bibnamefont {Rakher}}, \bibinfo {author}
  {\bibfnamefont {M.}~\bibnamefont {Korppi}}, \bibinfo {author} {\bibfnamefont
  {A.}~\bibnamefont {J{\"o}ckel}}, \ and\ \bibinfo {author} {\bibfnamefont
  {P.}~\bibnamefont {Treutlein}},\ }\href {\doibase 10.1103/PhysRevA.87.023816}
  {\bibfield  {journal} {\bibinfo  {journal} {Physical Review A}\ }\textbf
  {\bibinfo {volume} {87}},\ \bibinfo {pages} {023816} (\bibinfo {year}
  {2013})}\BibitemShut {NoStop}%
\bibitem [{\citenamefont {Bennett}\ \emph {et~al.}(2014)\citenamefont
  {Bennett}, \citenamefont {Madsen}, \citenamefont {Baker}, \citenamefont
  {Rubinsztein-Dunlop},\ and\ \citenamefont {Bowen}}]{Bennett:2014ui}%
  \BibitemOpen
  \bibfield  {author} {\bibinfo {author} {\bibfnamefont {J.~S.}\ \bibnamefont
  {Bennett}}, \bibinfo {author} {\bibfnamefont {L.~S.}\ \bibnamefont {Madsen}},
  \bibinfo {author} {\bibfnamefont {M.}~\bibnamefont {Baker}}, \bibinfo
  {author} {\bibfnamefont {H.}~\bibnamefont {Rubinsztein-Dunlop}}, \ and\
  \bibinfo {author} {\bibfnamefont {W.~P.}\ \bibnamefont {Bowen}},\ }\href
  {http://arxiv.org/abs/1404.3445} {\bibfield  {journal} {\bibinfo  {journal}
  {arXiv.org}\ ,\ \bibinfo {pages} {1404.3445}} (\bibinfo {year}
  {2014})}\BibitemShut {NoStop}%
\bibitem [{\citenamefont {Hunger}\ \emph {et~al.}(2011)\citenamefont {Hunger},
  \citenamefont {Camerer}, \citenamefont {Korppi}, \citenamefont {J{\"o}ckel},
  \citenamefont {H{\"a}nsch},\ and\ \citenamefont {Treutlein}}]{Hunger:2011eo}%
  \BibitemOpen
  \bibfield  {author} {\bibinfo {author} {\bibfnamefont {D.}~\bibnamefont
  {Hunger}}, \bibinfo {author} {\bibfnamefont {S.}~\bibnamefont {Camerer}},
  \bibinfo {author} {\bibfnamefont {M.}~\bibnamefont {Korppi}}, \bibinfo
  {author} {\bibfnamefont {A.}~\bibnamefont {J{\"o}ckel}}, \bibinfo {author}
  {\bibfnamefont {T.~W.}\ \bibnamefont {H{\"a}nsch}}, \ and\ \bibinfo {author}
  {\bibfnamefont {P.}~\bibnamefont {Treutlein}},\ }\href {\doibase
  10.1016/j.crhy.2011.04.015} {\bibfield  {journal} {\bibinfo  {journal}
  {Comptes Rendus Physique}\ }\textbf {\bibinfo {volume} {12}},\ \bibinfo
  {pages} {871} (\bibinfo {year} {2011})}\BibitemShut {NoStop}%
\bibitem [{\citenamefont {Treutlein}\ \emph {et~al.}(4151)\citenamefont
  {Treutlein}, \citenamefont {Genes}, \citenamefont {Hammerer}, \citenamefont
  {Poggio},\ and\ \citenamefont {Rabl}}]{Treutlein:2012wa}%
  \BibitemOpen
  \bibfield  {author} {\bibinfo {author} {\bibfnamefont {P.}~\bibnamefont
  {Treutlein}}, \bibinfo {author} {\bibfnamefont {C.}~\bibnamefont {Genes}},
  \bibinfo {author} {\bibfnamefont {K.}~\bibnamefont {Hammerer}}, \bibinfo
  {author} {\bibfnamefont {M.}~\bibnamefont {Poggio}}, \ and\ \bibinfo {author}
  {\bibfnamefont {P.}~\bibnamefont {Rabl}},\ }in\ \href
  {http://arxiv.org/abs/1210.4151} {\emph {\bibinfo {booktitle}
  {Cavity-Optomechanics}}},\ \bibinfo {editor} {edited by\ \bibinfo {editor}
  {\bibfnamefont {F.}~\bibnamefont {Marquardt}}, \bibinfo {editor}
  {\bibfnamefont {M.}~\bibnamefont {Aspelmeyer}}, \ and\ \bibinfo {editor}
  {\bibfnamefont {T.}~\bibnamefont {Kippenberg}}}\ (\bibinfo  {publisher}
  {Springer-Verlag},\ \bibinfo {address} {Berlin},\ \bibinfo {year} {2014,
  preprint arXiv:1210.4151})\BibitemShut {NoStop}%
\bibitem [{\citenamefont {Millen}\ \emph {et~al.}(2014)\citenamefont {Millen},
  \citenamefont {Deesuwan}, \citenamefont {Barker},\ and\ \citenamefont
  {Anders}}]{Millen:2014dd}%
  \BibitemOpen
  \bibfield  {author} {\bibinfo {author} {\bibfnamefont {J.}~\bibnamefont
  {Millen}}, \bibinfo {author} {\bibfnamefont {T.}~\bibnamefont {Deesuwan}},
  \bibinfo {author} {\bibfnamefont {P.}~\bibnamefont {Barker}}, \ and\ \bibinfo
  {author} {\bibfnamefont {J.}~\bibnamefont {Anders}},\ }\href {\doibase
  10.1038/nnano.2014.82} {\bibfield  {journal} {\bibinfo  {journal} {Nature
  Nanotechnology}\ }\textbf {\bibinfo {volume} {9}},\ \bibinfo {pages} {425}
  (\bibinfo {year} {2014})}\BibitemShut {NoStop}%
\bibitem [{\citenamefont {Aspelmeyer}\ \emph {et~al.}(2012)\citenamefont
  {Aspelmeyer}, \citenamefont {Meystre},\ and\ \citenamefont
  {Schwab}}]{Aspelmeyer:2012fy}%
  \BibitemOpen
  \bibfield  {author} {\bibinfo {author} {\bibfnamefont {M.}~\bibnamefont
  {Aspelmeyer}}, \bibinfo {author} {\bibfnamefont {P.}~\bibnamefont {Meystre}},
  \ and\ \bibinfo {author} {\bibfnamefont {K.}~\bibnamefont {Schwab}},\ }\href
  {\doibase 10.1063/PT.3.1640} {\bibfield  {journal} {\bibinfo  {journal}
  {Physics Today}\ }\textbf {\bibinfo {volume} {65}},\ \bibinfo {pages} {29}
  (\bibinfo {year} {2012})}\BibitemShut {NoStop}%
\bibitem [{\citenamefont {O'Connell}\ \emph {et~al.}(2010)\citenamefont
  {O'Connell}, \citenamefont {Hofheinz}, \citenamefont {Ansmann}, \citenamefont
  {Bialczak}, \citenamefont {Lenander}, \citenamefont {Lucero}, \citenamefont
  {Neeley}, \citenamefont {Sank}, \citenamefont {Wang}, \citenamefont {Weides},
  \citenamefont {Wenner}, \citenamefont {Martinis},\ and\ \citenamefont
  {Cleland}}]{Oconnell:2010br}%
  \BibitemOpen
  \bibfield  {author} {\bibinfo {author} {\bibfnamefont {A.~D.}\ \bibnamefont
  {O'Connell}}, \bibinfo {author} {\bibfnamefont {M.}~\bibnamefont {Hofheinz}},
  \bibinfo {author} {\bibfnamefont {M.}~\bibnamefont {Ansmann}}, \bibinfo
  {author} {\bibfnamefont {R.~C.}\ \bibnamefont {Bialczak}}, \bibinfo {author}
  {\bibfnamefont {M.}~\bibnamefont {Lenander}}, \bibinfo {author}
  {\bibfnamefont {E.}~\bibnamefont {Lucero}}, \bibinfo {author} {\bibfnamefont
  {M.}~\bibnamefont {Neeley}}, \bibinfo {author} {\bibfnamefont
  {D.}~\bibnamefont {Sank}}, \bibinfo {author} {\bibfnamefont {H.}~\bibnamefont
  {Wang}}, \bibinfo {author} {\bibfnamefont {M.}~\bibnamefont {Weides}},
  \bibinfo {author} {\bibfnamefont {J.}~\bibnamefont {Wenner}}, \bibinfo
  {author} {\bibfnamefont {J.~M.}\ \bibnamefont {Martinis}}, \ and\ \bibinfo
  {author} {\bibfnamefont {A.~N.}\ \bibnamefont {Cleland}},\ }\href {\doibase
  10.1038/nature08967} {\bibfield  {journal} {\bibinfo  {journal} {Nature}\
  }\textbf {\bibinfo {volume} {464}},\ \bibinfo {pages} {697} (\bibinfo {year}
  {2010})}\BibitemShut {NoStop}%
\bibitem [{\citenamefont {Teufel}\ \emph {et~al.}(2011)\citenamefont {Teufel},
  \citenamefont {Donner}, \citenamefont {Li}, \citenamefont {Harlow},
  \citenamefont {Allman}, \citenamefont {Cicak}, \citenamefont {Sirois},
  \citenamefont {Whittaker}, \citenamefont {Lehnert},\ and\ \citenamefont
  {Simmonds}}]{Teufel:2011jg}%
  \BibitemOpen
  \bibfield  {author} {\bibinfo {author} {\bibfnamefont {J.~D.}\ \bibnamefont
  {Teufel}}, \bibinfo {author} {\bibfnamefont {T.}~\bibnamefont {Donner}},
  \bibinfo {author} {\bibfnamefont {D.}~\bibnamefont {Li}}, \bibinfo {author}
  {\bibfnamefont {J.~W.}\ \bibnamefont {Harlow}}, \bibinfo {author}
  {\bibfnamefont {M.~S.}\ \bibnamefont {Allman}}, \bibinfo {author}
  {\bibfnamefont {K.}~\bibnamefont {Cicak}}, \bibinfo {author} {\bibfnamefont
  {A.~J.}\ \bibnamefont {Sirois}}, \bibinfo {author} {\bibfnamefont {J.~D.}\
  \bibnamefont {Whittaker}}, \bibinfo {author} {\bibfnamefont {K.~W.}\
  \bibnamefont {Lehnert}}, \ and\ \bibinfo {author} {\bibfnamefont {R.~W.}\
  \bibnamefont {Simmonds}},\ }\href {\doibase 10.1038/nature10261} {\bibfield
  {journal} {\bibinfo  {journal} {Nature}\ }\textbf {\bibinfo {volume} {475}},\
  \bibinfo {pages} {359} (\bibinfo {year} {2011})}\BibitemShut {NoStop}%
\bibitem [{\citenamefont {Chan}\ \emph {et~al.}(2011)\citenamefont {Chan},
  \citenamefont {Alegre}, \citenamefont {Safavi-Naeini}, \citenamefont {Hill},
  \citenamefont {Krause}, \citenamefont {Gr{\"o}blacher}, \citenamefont
  {Aspelmeyer},\ and\ \citenamefont {Painter}}]{Chan:2011dy}%
  \BibitemOpen
  \bibfield  {author} {\bibinfo {author} {\bibfnamefont {J.}~\bibnamefont
  {Chan}}, \bibinfo {author} {\bibfnamefont {T.~P.~M.}\ \bibnamefont {Alegre}},
  \bibinfo {author} {\bibfnamefont {A.~H.}\ \bibnamefont {Safavi-Naeini}},
  \bibinfo {author} {\bibfnamefont {J.~T.}\ \bibnamefont {Hill}}, \bibinfo
  {author} {\bibfnamefont {A.}~\bibnamefont {Krause}}, \bibinfo {author}
  {\bibfnamefont {S.}~\bibnamefont {Gr{\"o}blacher}}, \bibinfo {author}
  {\bibfnamefont {M.}~\bibnamefont {Aspelmeyer}}, \ and\ \bibinfo {author}
  {\bibfnamefont {O.}~\bibnamefont {Painter}},\ }\href {\doibase
  10.1038/nature10461} {\bibfield  {journal} {\bibinfo  {journal} {Nature}\
  }\textbf {\bibinfo {volume} {478}},\ \bibinfo {pages} {89} (\bibinfo {year}
  {2011})}\BibitemShut {NoStop}%
\bibitem [{\citenamefont {Verhagen}\ \emph {et~al.}(2013)\citenamefont
  {Verhagen}, \citenamefont {Deleglise}, \citenamefont {Weis}, \citenamefont
  {Schliesser},\ and\ \citenamefont {Kippenberg}}]{Verhagen:2013ei}%
  \BibitemOpen
  \bibfield  {author} {\bibinfo {author} {\bibfnamefont {E.}~\bibnamefont
  {Verhagen}}, \bibinfo {author} {\bibfnamefont {S.}~\bibnamefont {Deleglise}},
  \bibinfo {author} {\bibfnamefont {S.}~\bibnamefont {Weis}}, \bibinfo {author}
  {\bibfnamefont {A.}~\bibnamefont {Schliesser}}, \ and\ \bibinfo {author}
  {\bibfnamefont {T.~J.}\ \bibnamefont {Kippenberg}},\ }\href {\doibase
  10.1038/nature10787} {\bibfield  {journal} {\bibinfo  {journal} {Nature}\
  }\textbf {\bibinfo {volume} {482}},\ \bibinfo {pages} {63} (\bibinfo {year}
  {2013})}\BibitemShut {NoStop}%
\bibitem [{\citenamefont {Aspelmeyer}\ \emph {et~al.}(2013)\citenamefont
  {Aspelmeyer}, \citenamefont {Kippenberg},\ and\ \citenamefont
  {Marquardt}}]{Aspelmeyer:2013vr}%
  \BibitemOpen
  \bibfield  {author} {\bibinfo {author} {\bibfnamefont {M.}~\bibnamefont
  {Aspelmeyer}}, \bibinfo {author} {\bibfnamefont {T.~J.}\ \bibnamefont
  {Kippenberg}}, \ and\ \bibinfo {author} {\bibfnamefont {F.}~\bibnamefont
  {Marquardt}},\ }\href {http://arxiv.org/abs/1303.0733} {\bibfield  {journal}
  {\bibinfo  {journal} {arXiv}\ ,\ \bibinfo {pages} {1303.0733}} (\bibinfo
  {year} {2013})},\ \bibinfo {note} {(to appear in \textit{Reviews of Modern
  Physics})}\BibitemShut {NoStop}%
\bibitem [{\citenamefont {Wang}\ \emph {et~al.}(2006)\citenamefont {Wang},
  \citenamefont {Eardley}, \citenamefont {Knappe}, \citenamefont {Moreland},
  \citenamefont {Hollberg},\ and\ \citenamefont {Kitching}}]{Wang:2006dd}%
  \BibitemOpen
  \bibfield  {author} {\bibinfo {author} {\bibfnamefont {Y.-J.}\ \bibnamefont
  {Wang}}, \bibinfo {author} {\bibfnamefont {M.}~\bibnamefont {Eardley}},
  \bibinfo {author} {\bibfnamefont {S.}~\bibnamefont {Knappe}}, \bibinfo
  {author} {\bibfnamefont {J.}~\bibnamefont {Moreland}}, \bibinfo {author}
  {\bibfnamefont {L.}~\bibnamefont {Hollberg}}, \ and\ \bibinfo {author}
  {\bibfnamefont {J.}~\bibnamefont {Kitching}},\ }\href {\doibase
  10.1103/PhysRevLett.97.227602} {\bibfield  {journal} {\bibinfo  {journal}
  {Physical Review Letters}\ }\textbf {\bibinfo {volume} {97}},\ \bibinfo
  {pages} {227602} (\bibinfo {year} {2006})}\BibitemShut {NoStop}%
\bibitem [{\citenamefont {Hunger}\ \emph {et~al.}(2010)\citenamefont {Hunger},
  \citenamefont {Camerer}, \citenamefont {H{\"a}nsch}, \citenamefont
  {K{\"o}nig}, \citenamefont {Kotthaus}, \citenamefont {Reichel},\ and\
  \citenamefont {Treutlein}}]{Hunger:2010fr}%
  \BibitemOpen
  \bibfield  {author} {\bibinfo {author} {\bibfnamefont {D.}~\bibnamefont
  {Hunger}}, \bibinfo {author} {\bibfnamefont {S.}~\bibnamefont {Camerer}},
  \bibinfo {author} {\bibfnamefont {T.~W.}\ \bibnamefont {H{\"a}nsch}},
  \bibinfo {author} {\bibfnamefont {D.}~\bibnamefont {K{\"o}nig}}, \bibinfo
  {author} {\bibfnamefont {J.~P.}\ \bibnamefont {Kotthaus}}, \bibinfo {author}
  {\bibfnamefont {J.}~\bibnamefont {Reichel}}, \ and\ \bibinfo {author}
  {\bibfnamefont {P.}~\bibnamefont {Treutlein}},\ }\href {\doibase
  10.1103/PhysRevLett.104.143002} {\bibfield  {journal} {\bibinfo  {journal}
  {Physical Review Letters}\ }\textbf {\bibinfo {volume} {104}},\ \bibinfo
  {pages} {143002} (\bibinfo {year} {2010})}\BibitemShut {NoStop}%
\bibitem [{\citenamefont {Degen}\ \emph {et~al.}(2009)\citenamefont {Degen},
  \citenamefont {Poggio}, \citenamefont {Mamin}, \citenamefont {Rettner},\ and\
  \citenamefont {Rugar}}]{Degen:2009kd}%
  \BibitemOpen
  \bibfield  {author} {\bibinfo {author} {\bibfnamefont {C.~L.}\ \bibnamefont
  {Degen}}, \bibinfo {author} {\bibfnamefont {M.}~\bibnamefont {Poggio}},
  \bibinfo {author} {\bibfnamefont {H.~J.}\ \bibnamefont {Mamin}}, \bibinfo
  {author} {\bibfnamefont {C.~T.}\ \bibnamefont {Rettner}}, \ and\ \bibinfo
  {author} {\bibfnamefont {D.}~\bibnamefont {Rugar}},\ }\href {\doibase
  10.1073/pnas.0812068106} {\bibfield  {journal} {\bibinfo  {journal}
  {Proceedings of the National Academy of Sciences}\ }\textbf {\bibinfo
  {volume} {106}},\ \bibinfo {pages} {1313} (\bibinfo {year}
  {2009})}\BibitemShut {NoStop}%
\bibitem [{\citenamefont {Arcizet}\ \emph {et~al.}(2011)\citenamefont
  {Arcizet}, \citenamefont {Jacques}, \citenamefont {Siria}, \citenamefont
  {Poncharal}, \citenamefont {Vincent},\ and\ \citenamefont
  {Seidelin}}]{Arcizet:2011cg}%
  \BibitemOpen
  \bibfield  {author} {\bibinfo {author} {\bibfnamefont {O.}~\bibnamefont
  {Arcizet}}, \bibinfo {author} {\bibfnamefont {V.}~\bibnamefont {Jacques}},
  \bibinfo {author} {\bibfnamefont {A.}~\bibnamefont {Siria}}, \bibinfo
  {author} {\bibfnamefont {P.}~\bibnamefont {Poncharal}}, \bibinfo {author}
  {\bibfnamefont {P.}~\bibnamefont {Vincent}}, \ and\ \bibinfo {author}
  {\bibfnamefont {S.}~\bibnamefont {Seidelin}},\ }\href {\doibase
  10.1038/nphys2070} {\bibfield  {journal} {\bibinfo  {journal} {Nature
  Physics}\ }\textbf {\bibinfo {volume} {7}},\ \bibinfo {pages} {879} (\bibinfo
  {year} {2011})}\BibitemShut {NoStop}%
\bibitem [{\citenamefont {Kolkowitz}\ \emph {et~al.}(2012)\citenamefont
  {Kolkowitz}, \citenamefont {Bleszynski~Jayich}, \citenamefont
  {Unterreithmeier}, \citenamefont {Bennett}, \citenamefont {Rabl},
  \citenamefont {Harris},\ and\ \citenamefont {Lukin}}]{Kolkowitz:2012iw}%
  \BibitemOpen
  \bibfield  {author} {\bibinfo {author} {\bibfnamefont {S.}~\bibnamefont
  {Kolkowitz}}, \bibinfo {author} {\bibfnamefont {A.~C.}\ \bibnamefont
  {Bleszynski~Jayich}}, \bibinfo {author} {\bibfnamefont {Q.~P.}\ \bibnamefont
  {Unterreithmeier}}, \bibinfo {author} {\bibfnamefont {S.~D.}\ \bibnamefont
  {Bennett}}, \bibinfo {author} {\bibfnamefont {P.}~\bibnamefont {Rabl}},
  \bibinfo {author} {\bibfnamefont {J.~G.~E.}\ \bibnamefont {Harris}}, \ and\
  \bibinfo {author} {\bibfnamefont {M.~D.}\ \bibnamefont {Lukin}},\ }\href
  {\doibase 10.1126/science.1216821} {\bibfield  {journal} {\bibinfo  {journal}
  {Science}\ }\textbf {\bibinfo {volume} {335}},\ \bibinfo {pages} {1603}
  (\bibinfo {year} {2012})}\BibitemShut {NoStop}%
\bibitem [{\citenamefont {Yeo}\ \emph {et~al.}(2013)\citenamefont {Yeo},
  \citenamefont {de~Assis}, \citenamefont {Gloppe}, \citenamefont
  {Dupont-Ferrier}, \citenamefont {Verlot}, \citenamefont {Malik},
  \citenamefont {Dupuy}, \citenamefont {Claudon}, \citenamefont {G{\'e}rard},
  \citenamefont {Auffeves}, \citenamefont {Nogues}, \citenamefont {Seidelin},
  \citenamefont {Poizat}, \citenamefont {Arcizet},\ and\ \citenamefont
  {Richard}}]{Yeo:2013ja}%
  \BibitemOpen
  \bibfield  {author} {\bibinfo {author} {\bibfnamefont {I.}~\bibnamefont
  {Yeo}}, \bibinfo {author} {\bibfnamefont {P.-L.}\ \bibnamefont {de~Assis}},
  \bibinfo {author} {\bibfnamefont {A.}~\bibnamefont {Gloppe}}, \bibinfo
  {author} {\bibfnamefont {E.}~\bibnamefont {Dupont-Ferrier}}, \bibinfo
  {author} {\bibfnamefont {P.}~\bibnamefont {Verlot}}, \bibinfo {author}
  {\bibfnamefont {N.~S.}\ \bibnamefont {Malik}}, \bibinfo {author}
  {\bibfnamefont {E.}~\bibnamefont {Dupuy}}, \bibinfo {author} {\bibfnamefont
  {J.}~\bibnamefont {Claudon}}, \bibinfo {author} {\bibfnamefont {J.-M.}\
  \bibnamefont {G{\'e}rard}}, \bibinfo {author} {\bibfnamefont
  {A.}~\bibnamefont {Auffeves}}, \bibinfo {author} {\bibfnamefont
  {G.}~\bibnamefont {Nogues}}, \bibinfo {author} {\bibfnamefont
  {S.}~\bibnamefont {Seidelin}}, \bibinfo {author} {\bibfnamefont {J.-P.}\
  \bibnamefont {Poizat}}, \bibinfo {author} {\bibfnamefont {O.}~\bibnamefont
  {Arcizet}}, \ and\ \bibinfo {author} {\bibfnamefont {M.}~\bibnamefont
  {Richard}},\ }\href {\doibase 10.1038/nnano.2013.274} {\bibfield  {journal}
  {\bibinfo  {journal} {Nature Nanotechnology}\ }\textbf {\bibinfo {volume}
  {9}},\ \bibinfo {pages} {106} (\bibinfo {year} {2013})}\BibitemShut {NoStop}%
\bibitem [{\citenamefont {Pirkkalainen}\ \emph {et~al.}(2013)\citenamefont
  {Pirkkalainen}, \citenamefont {Cho}, \citenamefont {Li}, \citenamefont
  {Paraoanu}, \citenamefont {Hakonen},\ and\ \citenamefont
  {Sillanp{\"a}{\"a}}}]{Pirkkalainen:2013gh}%
  \BibitemOpen
  \bibfield  {author} {\bibinfo {author} {\bibfnamefont {J.~M.}\ \bibnamefont
  {Pirkkalainen}}, \bibinfo {author} {\bibfnamefont {S.~U.}\ \bibnamefont
  {Cho}}, \bibinfo {author} {\bibfnamefont {J.}~\bibnamefont {Li}}, \bibinfo
  {author} {\bibfnamefont {G.~S.}\ \bibnamefont {Paraoanu}}, \bibinfo {author}
  {\bibfnamefont {P.~J.}\ \bibnamefont {Hakonen}}, \ and\ \bibinfo {author}
  {\bibfnamefont {M.~A.}\ \bibnamefont {Sillanp{\"a}{\"a}}},\ }\href {\doibase
  10.1038/nature11821} {\bibfield  {journal} {\bibinfo  {journal} {Nature}\
  }\textbf {\bibinfo {volume} {494}},\ \bibinfo {pages} {211} (\bibinfo {year}
  {2013})}\BibitemShut {NoStop}%
\bibitem [{\citenamefont {Weidem{\"u}ller}\ and\ \citenamefont
  {Zimmermann}(2009)}]{Weidemueller:2009}%
  \BibitemOpen
  \bibinfo {editor} {\bibfnamefont {M.}~\bibnamefont {Weidem{\"u}ller}}\ and\
  \bibinfo {editor} {\bibfnamefont {C.}~\bibnamefont {Zimmermann}},\ eds.,\
  \href@noop {} {\emph {\bibinfo {title} {Cold Atoms and Molecules}}}\
  (\bibinfo  {publisher} {Wiley-VCH},\ \bibinfo {address} {Berlin},\ \bibinfo
  {year} {2009})\BibitemShut {NoStop}%
\bibitem [{\citenamefont {Genes}\ \emph {et~al.}(2011)\citenamefont {Genes},
  \citenamefont {Ritsch}, \citenamefont {Drewsen},\ and\ \citenamefont
  {Dantan}}]{Genes:2011jd}%
  \BibitemOpen
  \bibfield  {author} {\bibinfo {author} {\bibfnamefont {C.}~\bibnamefont
  {Genes}}, \bibinfo {author} {\bibfnamefont {H.}~\bibnamefont {Ritsch}},
  \bibinfo {author} {\bibfnamefont {M.}~\bibnamefont {Drewsen}}, \ and\
  \bibinfo {author} {\bibfnamefont {A.}~\bibnamefont {Dantan}},\ }\href
  {\doibase 10.1103/PhysRevA.84.051801} {\bibfield  {journal} {\bibinfo
  {journal} {Physical Review A}\ }\textbf {\bibinfo {volume} {84}},\ \bibinfo
  {pages} {051801} (\bibinfo {year} {2011})}\BibitemShut {NoStop}%
\bibitem [{\citenamefont {Hammerer}\ \emph {et~al.}(2009)\citenamefont
  {Hammerer}, \citenamefont {Aspelmeyer}, \citenamefont {Polzik},\ and\
  \citenamefont {Zoller}}]{Hammerer:2009gk}%
  \BibitemOpen
  \bibfield  {author} {\bibinfo {author} {\bibfnamefont {K.}~\bibnamefont
  {Hammerer}}, \bibinfo {author} {\bibfnamefont {M.}~\bibnamefont
  {Aspelmeyer}}, \bibinfo {author} {\bibfnamefont {E.}~\bibnamefont {Polzik}},
  \ and\ \bibinfo {author} {\bibfnamefont {P.}~\bibnamefont {Zoller}},\ }\href
  {\doibase 10.1103/PhysRevLett.102.020501} {\bibfield  {journal} {\bibinfo
  {journal} {Physical Review Letters}\ }\textbf {\bibinfo {volume} {102}},\
  \bibinfo {pages} {020501} (\bibinfo {year} {2009})}\BibitemShut {NoStop}%
\bibitem [{\citenamefont {Carmele}\ \emph {et~al.}(2013)\citenamefont
  {Carmele}, \citenamefont {Vogell}, \citenamefont {Stannigel},\ and\
  \citenamefont {Zoller}}]{Carmele:2013vf}%
  \BibitemOpen
  \bibfield  {author} {\bibinfo {author} {\bibfnamefont {A.}~\bibnamefont
  {Carmele}}, \bibinfo {author} {\bibfnamefont {B.}~\bibnamefont {Vogell}},
  \bibinfo {author} {\bibfnamefont {K.}~\bibnamefont {Stannigel}}, \ and\
  \bibinfo {author} {\bibfnamefont {P.}~\bibnamefont {Zoller}},\ }\href
  {http://arxiv.org/abs/1312.6551} {\bibfield  {journal} {\bibinfo  {journal}
  {arXiv}\ ,\ \bibinfo {pages} {1312.6551}} (\bibinfo {year}
  {2013})}\BibitemShut {NoStop}%
\bibitem [{\citenamefont {Willitsch}(2012)}]{Willitsch:2012gx}%
  \BibitemOpen
  \bibfield  {author} {\bibinfo {author} {\bibfnamefont {S.}~\bibnamefont
  {Willitsch}},\ }\href {\doibase 10.1080/0144235X.2012.667221} {\bibfield
  {journal} {\bibinfo  {journal} {International Reviews in Physical Chemistry}\
  }\textbf {\bibinfo {volume} {31}},\ \bibinfo {pages} {175} (\bibinfo {year}
  {2012})}\BibitemShut {NoStop}%
\bibitem [{\citenamefont {Jayich}\ \emph {et~al.}(2008)\citenamefont {Jayich},
  \citenamefont {Sankey}, \citenamefont {Zwickl}, \citenamefont {Yang},
  \citenamefont {Thompson}, \citenamefont {Girvin}, \citenamefont {Clerk},
  \citenamefont {Marquardt},\ and\ \citenamefont {Harris}}]{Jayich:2008iz}%
  \BibitemOpen
  \bibfield  {author} {\bibinfo {author} {\bibfnamefont {A.~M.}\ \bibnamefont
  {Jayich}}, \bibinfo {author} {\bibfnamefont {J.~C.}\ \bibnamefont {Sankey}},
  \bibinfo {author} {\bibfnamefont {B.~M.}\ \bibnamefont {Zwickl}}, \bibinfo
  {author} {\bibfnamefont {C.}~\bibnamefont {Yang}}, \bibinfo {author}
  {\bibfnamefont {J.~D.}\ \bibnamefont {Thompson}}, \bibinfo {author}
  {\bibfnamefont {S.~M.}\ \bibnamefont {Girvin}}, \bibinfo {author}
  {\bibfnamefont {A.~A.}\ \bibnamefont {Clerk}}, \bibinfo {author}
  {\bibfnamefont {F.}~\bibnamefont {Marquardt}}, \ and\ \bibinfo {author}
  {\bibfnamefont {J.~G.~E.}\ \bibnamefont {Harris}},\ }\href {\doibase
  10.1088/1367-2630/10/9/095008} {\bibfield  {journal} {\bibinfo  {journal}
  {New Journal of Physics}\ }\textbf {\bibinfo {volume} {10}},\ \bibinfo
  {pages} {095008} (\bibinfo {year} {2008})}\BibitemShut {NoStop}%
\bibitem [{\citenamefont {Elste}\ \emph {et~al.}(2009)\citenamefont {Elste},
  \citenamefont {Girvin},\ and\ \citenamefont {Clerk}}]{Elste:2009hf}%
  \BibitemOpen
  \bibfield  {author} {\bibinfo {author} {\bibfnamefont {F.}~\bibnamefont
  {Elste}}, \bibinfo {author} {\bibfnamefont {S.~M.}\ \bibnamefont {Girvin}}, \
  and\ \bibinfo {author} {\bibfnamefont {A.~A.}\ \bibnamefont {Clerk}},\ }\href
  {\doibase 10.1103/PhysRevLett.102.207209} {\bibfield  {journal} {\bibinfo
  {journal} {Physical Review Letters}\ }\textbf {\bibinfo {volume} {102}},\
  \bibinfo {pages} {207209} (\bibinfo {year} {2009})}\BibitemShut {NoStop}%
\bibitem [{\citenamefont {Weiss}\ \emph {et~al.}(2013)\citenamefont {Weiss},
  \citenamefont {Bruder},\ and\ \citenamefont {Nunnenkamp}}]{Weiss:2013gr}%
  \BibitemOpen
  \bibfield  {author} {\bibinfo {author} {\bibfnamefont {T.}~\bibnamefont
  {Weiss}}, \bibinfo {author} {\bibfnamefont {C.}~\bibnamefont {Bruder}}, \
  and\ \bibinfo {author} {\bibfnamefont {A.}~\bibnamefont {Nunnenkamp}},\
  }\href {\doibase 10.1088/1367-2630/15/4/045017} {\bibfield  {journal}
  {\bibinfo  {journal} {New Journal of Physics}\ }\textbf {\bibinfo {volume}
  {15}},\ \bibinfo {pages} {045017} (\bibinfo {year} {2013})}\BibitemShut
  {NoStop}%
\bibitem [{\citenamefont {J{\"o}ckel}\ \emph {et~al.}(2011)\citenamefont
  {J{\"o}ckel}, \citenamefont {Rakher}, \citenamefont {Korppi}, \citenamefont
  {Camerer}, \citenamefont {Hunger}, \citenamefont {Mader},\ and\ \citenamefont
  {Treutlein}}]{Jockel:2011ex}%
  \BibitemOpen
  \bibfield  {author} {\bibinfo {author} {\bibfnamefont {A.}~\bibnamefont
  {J{\"o}ckel}}, \bibinfo {author} {\bibfnamefont {M.~T.}\ \bibnamefont
  {Rakher}}, \bibinfo {author} {\bibfnamefont {M.}~\bibnamefont {Korppi}},
  \bibinfo {author} {\bibfnamefont {S.}~\bibnamefont {Camerer}}, \bibinfo
  {author} {\bibfnamefont {D.}~\bibnamefont {Hunger}}, \bibinfo {author}
  {\bibfnamefont {M.}~\bibnamefont {Mader}}, \ and\ \bibinfo {author}
  {\bibfnamefont {P.}~\bibnamefont {Treutlein}},\ }\href {\doibase
  10.1063/1.3646914} {\bibfield  {journal} {\bibinfo  {journal} {Appl. Phys.
  Lett.}\ }\textbf {\bibinfo {volume} {99}},\ \bibinfo {pages} {143109}
  (\bibinfo {year} {2011})}\BibitemShut {NoStop}%
\bibitem [{\citenamefont {Grimm}\ \emph {et~al.}(2000)\citenamefont {Grimm},
  \citenamefont {Weidem{\"u}ller},\ and\ \citenamefont
  {Ovchinnikov}}]{Grimm:2000wx}%
  \BibitemOpen
  \bibfield  {author} {\bibinfo {author} {\bibfnamefont {R.}~\bibnamefont
  {Grimm}}, \bibinfo {author} {\bibfnamefont {M.}~\bibnamefont
  {Weidem{\"u}ller}}, \ and\ \bibinfo {author} {\bibfnamefont {Y.~B.}\
  \bibnamefont {Ovchinnikov}},\ }\href {\doibase 10.1016/S1049-250X(08)60186-X}
  {\bibfield  {journal} {\bibinfo  {journal} {Adv. At. Mol. Opt. Phys}\
  }\textbf {\bibinfo {volume} {42}},\ \bibinfo {pages} {95} (\bibinfo {year}
  {2000})}\BibitemShut {NoStop}%
\bibitem [{\citenamefont {Morrow}\ \emph {et~al.}(2002)\citenamefont {Morrow},
  \citenamefont {Dutta},\ and\ \citenamefont {Raithel}}]{Morrow:2002vp}%
  \BibitemOpen
  \bibfield  {author} {\bibinfo {author} {\bibfnamefont {N.}~\bibnamefont
  {Morrow}}, \bibinfo {author} {\bibfnamefont {S.}~\bibnamefont {Dutta}}, \
  and\ \bibinfo {author} {\bibfnamefont {G.}~\bibnamefont {Raithel}},\ }\href
  {\doibase 10.1103/PhysRevLett.88.093003} {\bibfield  {journal} {\bibinfo
  {journal} {Physical Review Letters}\ }\textbf {\bibinfo {volume} {88}},\
  \bibinfo {pages} {093003} (\bibinfo {year} {2002})}\BibitemShut {NoStop}%
\bibitem [{\citenamefont {Chakram}\ \emph {et~al.}(2014)\citenamefont
  {Chakram}, \citenamefont {Patil}, \citenamefont {Chang},\ and\ \citenamefont
  {Vengalattore}}]{Chakram:2014ea}%
  \BibitemOpen
  \bibfield  {author} {\bibinfo {author} {\bibfnamefont {S.}~\bibnamefont
  {Chakram}}, \bibinfo {author} {\bibfnamefont {Y.~S.}\ \bibnamefont {Patil}},
  \bibinfo {author} {\bibfnamefont {L.}~\bibnamefont {Chang}}, \ and\ \bibinfo
  {author} {\bibfnamefont {M.}~\bibnamefont {Vengalattore}},\ }\href {\doibase
  10.1103/PhysRevLett.112.127201} {\bibfield  {journal} {\bibinfo  {journal}
  {Physical Review Letters}\ }\textbf {\bibinfo {volume} {112}},\ \bibinfo
  {pages} {127201} (\bibinfo {year} {2014})}\BibitemShut {NoStop}%
\bibitem [{\citenamefont {Purdy}\ \emph {et~al.}(2013)\citenamefont {Purdy},
  \citenamefont {Peterson},\ and\ \citenamefont {Regal}}]{Purdy:2013cb}%
  \BibitemOpen
  \bibfield  {author} {\bibinfo {author} {\bibfnamefont {T.~P.}\ \bibnamefont
  {Purdy}}, \bibinfo {author} {\bibfnamefont {R.~W.}\ \bibnamefont {Peterson}},
  \ and\ \bibinfo {author} {\bibfnamefont {C.~A.}\ \bibnamefont {Regal}},\
  }\href {\doibase 10.1126/science.1231282} {\bibfield  {journal} {\bibinfo
  {journal} {Science}\ }\textbf {\bibinfo {volume} {339}},\ \bibinfo {pages}
  {801} (\bibinfo {year} {2013})}\BibitemShut {NoStop}%
\bibitem [{\citenamefont {Kerman}\ \emph {et~al.}(2000)\citenamefont {Kerman},
  \citenamefont {Vuletic}, \citenamefont {Chin},\ and\ \citenamefont
  {Chu}}]{Kerman:2000th}%
  \BibitemOpen
  \bibfield  {author} {\bibinfo {author} {\bibfnamefont {A.}~\bibnamefont
  {Kerman}}, \bibinfo {author} {\bibfnamefont {V.}~\bibnamefont {Vuletic}},
  \bibinfo {author} {\bibfnamefont {C.}~\bibnamefont {Chin}}, \ and\ \bibinfo
  {author} {\bibfnamefont {S.}~\bibnamefont {Chu}},\ }\href {\doibase
  10.1103/PhysRevLett.84.439} {\bibfield  {journal} {\bibinfo  {journal}
  {Physical Review Letters}\ }\textbf {\bibinfo {volume} {84}},\ \bibinfo
  {pages} {439} (\bibinfo {year} {2000})}\BibitemShut {NoStop}%
\end{thebibliography}%

\end{document}